\documentclass[12pt]{article}
\usepackage{epsfig,amsfonts,amsthm}
\usepackage{amsmath,amssymb}
\usepackage[normalem]{ulem}

\usepackage{cancel}

\usepackage{color}
\usepackage{array}
\newcommand{\be}{\begin{equation}}
\newcommand{\ee}{\end{equation}}
\newcommand{\bea}{\begin{eqnarray}}
\newcommand{\eea}{\end{eqnarray}}

\newcommand{\doublet}[2]{ \left( \begin{array}{c}#1 \\ #2 \end{array}\right) }

\newcommand{\gev}{\mathrm{\;GeV}} 
\newcolumntype{C}[1]{>{\centering}m{#1}}

\def\lsim{\mathrel{\rlap{\lower4pt\hbox{\hskip1pt$\sim$}}
    \raise1pt\hbox{$<$}}}         %less than or approx. symbol
\def\gsim{\mathrel{\rlap{\lower4pt\hbox{\hskip1pt$\sim$}}
    \raise1pt\hbox{$>$}}}         %greater than or approx. symbol
    
\usepackage{amsmath}
\usepackage{amsfonts}
\usepackage{amssymb}
\usepackage{subfig}
\usepackage{wrapfig}
\usepackage{graphicx}

\def\beq{\begin{equation}}
\def\eeq{\end{equation}}
\def\bea{\begin{eqnarray}}
\def\eea{\end{eqnarray}}

\def\<{\left\langle}
\def\>{\right\rangle}

\newcommand{\bt}{\begin{tabular}}
\newcommand{\et}{\end{tabular}}

\newcommand{\Et}{\cancel{E}_{T}}

\usepackage{slashed}%slashed paketti antaa nelivektori kertaa gammamatriisi jutun "pee slashin"
\usepackage{amssymb} %tää antaa AMS symboleita joita ei näemmä voi muuten saada. esim nuoli oikealle jossa on kieltoviiva päällä \nrightarrow
\usepackage{float}%Tämä mahdollistaa kuvien laittamisen minipageen

\usepackage{tikz}
\usetikzlibrary{decorations.pathmorphing,decorations.markings}

\usepackage{graphicx}%Täl saa laitettuu kuvii. Esim. ps filei. Kun haluu laittaa .epsi tai .eps tiedostoja, niin laittaa eka \includegraphics{nimi.eps}. Sit kääntää pdflatex:lla ja kuva tulee. 

\tikzset{
photon/.style={decorate, decoration={snake,amplitude=2pt, segment length=5pt}, draw=black},
particle/.style={draw=black, postaction={decorate}, decoration={markings,mark=at position .5 with {\arrow[draw=black]{>}}}},
antiparticle/.style={draw=black, postaction={decorate}, decoration={markings,mark=at position .5 with {\arrowreversed[draw=black]{>}}}},
gluon/.style={decorate, draw=black, decoration={coil,amplitude=4pt, segment length=5pt}},
goldstone/.style={draw=green,postaction={decorate},decoration={markings,mark=at position .5 with {\arrow[draw=blue]{>}}}}
}

\usepackage[
top    = 3.5cm,
bottom = 3.5cm,
left   = 3.5cm,
right  = 3.5cm]{geometry}

\allowdisplaybreaks[2]
\begin{document}
\bibliographystyle{OurBibTeX}

\title{\hfill ~\\[-30mm]
                  \textbf{Observable Heavy Higgs Dark Matter 
                }        }
\date{}
\author{\\[-5mm]
Venus ~Keus\footnote{E-mail: {\tt Venus.Keus@helsinki.fi}} $^{1,2}$,\ 
Stephen ~F .~King\footnote{E-mail: {\tt King@soton.ac.uk}} $^{2}$,\ \\
Stefano ~Moretti\footnote{E-mail: {\tt S.Moretti@soton.ac.uk}}, $^{2,3}$\ 
Dorota ~Sokolowska \footnote{E-mail: {\tt Dorota.Sokolowska@fuw.edu.pl}} $^{4}$
\\ \\
  \emph{\small $^1$ Department of Physics and Helsinki Institute of Physics,}\\
 \emph{\small Gustaf Hallstromin katu 2, FIN-00014 University of Helsinki, Finland}\\
  \emph{\small $^2$ School of Physics and Astronomy, University of Southampton,}\\
  \emph{\small Southampton, SO17 1BJ, United Kingdom}\\
  \emph{\small  $^3$ Particle Physics Department, Rutherford Appleton Laboratory,}\\
 \emph{\small Chilton, Didcot, Oxon OX11 0QX, United Kingdom}\\
  \emph{\small  $^4$ University of Warsaw, Faculty of Physics, Pasteura 5,}\\
  \emph{\small 02-093 Warsaw, Poland}\\[4mm]}
\maketitle

\vspace*{-0.250truecm}
\begin{abstract}
\noindent
{Dark Matter (DM), arising from an Inert Higgs Doublet, may either be light, below the $W$ mass, or heavy, above about 525 GeV. While the light region may soon be excluded, the heavy region is known to be very difficult to probe with either Direct Detection (DD) experiments or the Large Hadron Collider (LHC).
We show that adding a second Inert Higgs Doublet helps to make the heavy DM region accessible to 
both DD and the LHC, by either increasing its couplings to the observed Higgs boson, or lowering its 
mass to $360 \gev \lesssim m_{DM}$, or both.} 
\end{abstract}
 
\thispagestyle{empty}
\vfill
\newpage
\setcounter{page}{1}

%%%%%%%%%%%%%%%%%%%%%%%%%%%%%%%%%%%%%%%%%%%%%%%%%%%%%%%%%%%%%%%%%%%%%%%%%%%%%%%%%%%%%%%%%%%%%%%%%%%%%%%%
\section{Introduction}

The long-awaited Higgs boson, with a mass of $m_h\approx 125$ GeV, was famously discovered in 2012 by the ATLAS and CMS experiments at the CERN Large Hadron Collider (LHC) \cite{Aad:2012tfa,Chatrchyan:2012ufa}. Although its properties are in accordance with the predictions of the Standard Model (SM), including Electro-Weak (EW) precision data, it remains an intruiging possibility that the observed Higgs boson, denoted here as $h$, may just be one member of an extended Higgs sector. A good motivation for the latter is the idea that it might provide a
candidate for Cold Dark Matter (CDM). 

Although the nature of Dark Matter (DM) is not yet known, according to the 
Standard Cosmological Lambda-CDM Model \cite{Ade:2015xua} it should be a
particle which is \textit{stable} on cosmological time scales, \textit{cold}, i.e., non-relativistic at the onset of galaxy formation, \textit{non-baryonic}, \textit{neutral} and \textit{weakly interacting}. Various candidates for such a state exist in the literature, the most well-studied being the Weakly Interacting Massive Particles (WIMPs) \cite{Jungman:1995df,Bertone:2004pz,Bergstrom:2000pn}, with masses between a few GeV and a few TeV. Any such WIMP candidate must be cosmologically 
stable, usually due to the conservation of a discrete symmetry,
and must freeze-out (i.e., drop out of thermal equilibrium) to yield the 
observed relic density \cite{Ade:2015xua}\footnote{Since the Planck 2015 results quotes various 
results for $\Omega_{DM} h^2$,
depending on which spectra such as TT, TE and EE are used, we prefer to use here the 
Planck 2013 result whose error encompasses all of them.}:
\begin{equation}\label{relic}
\Omega_{DM} h^2 = 0.1199 \pm 0.0027. 
\end{equation}

It is clear that the SM Higgs sector cannot provide a WIMP candidate, since its Higgs boson
is unstable. However, it was suggested some time ago that the Higgs sector could be extended by the addition of an extra doublet, which may not develop
a Vacuum Expectation Value (VEV), leaving a discrete $Z_2$ symmetry unbroken \cite{Deshpande:1977rw}.
Independently, it was later shown that an extra scalar doublet with zero VEV, odd under a discrete $Z_2$ symmetry, 
could yield monojets at hadron colliders while being constrained by DM considerations 
(the first time to our knowledge that any connection with hadron colliders or DM was made) \cite{King:1984vr}.
This possibility, which became known as the Inert Doublet Model (IDM),
has been studied extensively for the last few years (see, e.g., \cite{Ma:2006km,Barbieri:2006dq,LopezHonorez:2006gr}). 
Since the IDM involves {\em 1} Inert Doublet plus {\em 1} active Higgs Doublet, we shall also refer to it henceforth as the I(1+1)HDM. 

In the IDM, aka the I(1+1)HDM, one extra spin-zero $SU(2)_L$ doublet with the same quantum numbers as the SM Higgs doublet is introduced. One of the possible vacuum states in this model involves the first doublet acquiring a VEV, henceforth called the \textit{active doublet}, while the second doublet does not develop a VEV and is referred to as the \textit{inert doublet} since it does not take part in EW Symmetry Breaking (EWSB). Since this doublet does not couple to fermions and it is by construction the only $Z_2$-odd field in the model, it provides a stable DM candidate, namely the lightest state among scalar and pseudo-scalar $Z_2$-odd particles.

The I(1+1)HDM remains a viable model for a scalar DM candidate, being in agreement with current experimental constraints. As of now, there are two regions of DM masses where one can expect viable solutions: a low DM mass region, $53 \gev \lesssim m_{DM} \lesssim m_W$ and a heavy DM mass region, $m_{DM} \gtrsim 525 \gev$. The most recent experimental data, both from direct detection experiments and from the LHC, has reduced the viable parameter space in the low mass region \cite{Krawczyk:2013jta,Arhrib:2013ela}. In the heavy mass region, however, where the sensitivity of DM direct detection experiments decreases significantly with increasing DM mass, the DM candidate may escape possible detection in this model.

In a recent paper \cite{Keus:2014jha} we studied DM in a model with {\em 2} inert Higgs plus {\em 1} active Higgs doublet, which we referred to as the I(2+1)HDM. In particular we focused on the region of parameter
space of the I(2+1)HDM where the DM candidate, the lightest inert scalar, is in the light mass region $(m_{DM} \lesssim m_W)$. We found that the extended scalar sector can relax the exclusion limits from direct detection experiments, providing a viable DM candidate in a region of parameter space which 
would be excluded in the I(1+1)HDM. 
In this paper we study the heavy DM mass region of the I(2+1)HDM.
We show that heavy Higgs DM in this model becomes more readily observable as a result of either lowering the DM mass to $360 \gev \lesssim m_{DM}$, or increasing the DM-Higgs coupling, or both,
while always maintaining the DM relic density within the required region. 

The layout of the remainder of this paper is as follows. In section \ref{sec:model} we review the I(2+1)HDM and focus on a simplified version of the model based on a smaller number of parameters. In section \ref{sec:annihilation} we calculate the relic density in the I(2+1)HDM,
discussing the relevant DM annihilation scenarios, including the extended co-annhilating (pseudo-)scalar sector. 
Section \ref{sec:DM} will be focused on new features of the I(2+1)HDM with respect to the I(1+1)HDM in the context of DM phenomenology, including enhanced DM-Higgs couplings and 
the new mass region $360 \gev \lesssim m_{DM} \lesssim 525 \gev$ as well as on 
discussing the resulting improved prospects for direct detection. In section \ref{sec:LHC} we present 
heavy DM signals via Higgs mediation at the LHC in the I(2+1)HDM which look more promising than in the I(1+1)HDM. Finally in section \ref{conclusions} we conclude the paper.

\section{The I(2+1)HDM}\label{sec:model}

\subsection{The scalar potential}\label{construction}

It has been shown in \cite{Ivanov:2011ae} that an N-Higgs-Doublet Model potential symmetric under a group $G$ of phase rotations can be divided into two parts;
a phase invariant part, $V_0$, and a collection of extra terms ensuring the symmetry group $G$, $V_G$.

We construct our $Z_2$-symmetric 3-Higgs Doublet Model potential generated by the group
\be 
\label{generator}
g=  \mathrm{diag}\left(-1, -1, 1 \right). 
\ee
which is of the following form\footnote{Note that adding extra $Z_2$-respecting terms such as
$ 
(\phi_3^\dagger\phi_1)(\phi_2^\dagger\phi_3), 
(\phi_1^\dagger\phi_2)(\phi_3^\dagger\phi_3), 
(\phi_1^\dagger\phi_2)(\phi_1^\dagger\phi_1)$ and/or 
$(\phi_1^\dagger\phi_2)(\phi_2^\dagger\phi_2)
$
does not change the phenomenology of the model. The coefficients of these terms, therefore, have been set to zero for simplicity.}:
\bea
\label{V0-3HDM}
V_{I(2+1)HDM}&=&V_0+V_{Z_2} \\
V_0 &=& - \mu^2_{1} (\phi_1^\dagger \phi_1) -\mu^2_2 (\phi_2^\dagger \phi_2) - \mu^2_3(\phi_3^\dagger \phi_3) \nonumber\\
&&+ \lambda_{11} (\phi_1^\dagger \phi_1)^2+ \lambda_{22} (\phi_2^\dagger \phi_2)^2  + \lambda_{33} (\phi_3^\dagger \phi_3)^2 \nonumber\\
&& + \lambda_{12}  (\phi_1^\dagger \phi_1)(\phi_2^\dagger \phi_2)
 + \lambda_{23}  (\phi_2^\dagger \phi_2)(\phi_3^\dagger \phi_3) + \lambda_{31} (\phi_3^\dagger \phi_3)(\phi_1^\dagger \phi_1) \nonumber\\
&& + \lambda'_{12} (\phi_1^\dagger \phi_2)(\phi_2^\dagger \phi_1) 
 + \lambda'_{23} (\phi_2^\dagger \phi_3)(\phi_3^\dagger \phi_2) + \lambda'_{31} (\phi_3^\dagger \phi_1)(\phi_1^\dagger \phi_3).  \nonumber\\
 V_{Z_2} &=& -\mu^2_{12}(\phi_1^\dagger\phi_2)+  \lambda_{1}(\phi_1^\dagger\phi_2)^2 + \lambda_2(\phi_2^\dagger\phi_3)^2 + \lambda_3(\phi_3^\dagger\phi_1)^2  + h.c. \nonumber
\eea

We shall not consider CP-violation in this paper, therefore, we require all parameters of the potential to be real.

The doublets are defined as
\be 
\phi_1= \doublet{$\begin{scriptsize}$ \phi^+_1 $\end{scriptsize}$}{\frac{H^0_1+iA^0_1}{\sqrt{2}}},\quad 
\phi_2= \doublet{$\begin{scriptsize}$ \phi^+_2 $\end{scriptsize}$}{\frac{H^0_2+iA^0_2}{\sqrt{2}}}, \quad 
\phi_3= \doublet{$\begin{scriptsize}$ G^+ $\end{scriptsize}$}{\frac{v+h+iG^0}{\sqrt{2}}}, 
\label{explicit-fields}
\ee
where $\phi_1$ and $\phi_2$ are the two inert doublets and $\phi_3$ is the one active doublet which plays the role of the SM Higgs doublet, with $h$ being the SM-Higgs boson and $G^\pm,~ G^0$ are the would-be Goldstone bosons.

We assign $Z_2$ charges to each doublet according to the $Z_2$ generator in Eq.(\ref{generator}): odd-$Z_2$ charge to the inert doublets, $\phi_1$ and $\phi_2$, and even-$Z_2$ charge to the active doublet, $\phi_3$. It is clear that the symmetry of the potential is respected by the vacuum alignment $(0,0,\frac{v}{\sqrt{2}})$.
The neutral fields from the inert doublets could then in principle be DM candidates. These neutral fields are stabilised from decaying into SM particles as a result of the conserved $Z_2$ symmetry of the potential after EWSB.

To make sure that the entire Lagrangian and not only the scalar potential is $Z_2$ symmetric, we assign an even $Z_2$ parity to all SM particles, identical to the $Z_2$ parity of the only doublet that couples to them, i.e., the active doublet $\phi_3$. With this parity assignment Flavour Changing Neutral Currents (FCNCs) are avoided as the extra doublets are forbidden to couple to fermions by $Z_2$ conservation.

The Yukawa Lagrangian of the model is identical to the SM Yukawa Lagrangian, with $\phi_3$ playing the role of the SM Higgs doublet:
\bea 
\mathcal{L}_{Yukawa} &=& \Gamma^u_{mn} \bar{q}_{m,L} \tilde{\phi}_3 u_{n,R} + \Gamma^d_{mn} \bar{q}_{m,L} \phi_3 d_{n,R} \nonumber\\
&& +  \Gamma^e_{mn} \bar{l}_{m,L} \phi_3 e_{n,R} + \Gamma^{\nu}_{mn} \bar{l}_{m,L} \tilde{\phi}_3 {\nu}_{n,R} + h.c.  
\eea

\subsection{Mass eigenstates}

The minimum of the potential sits at the point $(0,0,\frac{v}{\sqrt{2}})$ with
$
v^2= \frac{\mu^2_3}{\lambda_{33}} .
$

\vspace{5mm}
\noindent The mass spectrum of the scalar particles are as follows.
\begin{itemize}
\item \textbf{The fields from the active doublet}\\
The fields from the third doublet, $G^0,G^\pm,h$, which play the role of the SM Higgs doublet fields have squared masses:
\bea 
&& m^2_{G^0}= m^2_{G^\pm}=0 \nonumber\\
&& m^2_{h}= 2\mu_3^2 
\eea

\item \textbf{The CP-even neutral inert fields}\\
The pair of inert neutral scalar gauge eigenstates, $H^0_{1},H^0_{2}$, which are rotated by
\be 
R_{\theta_h}= 
\left( \begin{array}{cc}
\cos \theta_h & \sin \theta_h \\
-\sin \theta_h & \cos \theta_h\\
\end{array} \right), \qquad \mbox{with} \quad \tan 2\theta_h = \frac{2\mu^2_{12}}{\mu^2_1 -\Lambda_{\phi_1} - \mu^2_2 + \Lambda_{\phi_2}}  \nonumber
\ee
into the mass eigenstates, $H_1, H_2$, have squared masses: 
\bea
&& m^2_{H_1}=  (-\mu^2_1 + \Lambda_{\phi_1})\cos^2\theta_h + (- \mu^2_2 + \Lambda_{\phi_2}) \sin^2\theta_h -2\mu^2_{12} \sin\theta_h \cos\theta_h \nonumber\\
&& m^2_{H_2}=  (-\mu^2_1 + \Lambda_{\phi_1})\sin^2\theta_h + (- \mu^2_2 + \Lambda_{\phi_2}) \cos^2\theta_h + 2\mu^2_{12} \sin\theta_h \cos\theta_h \nonumber\\
&& \quad \mbox{where} \quad \Lambda_{\phi_1}= \frac{1}{2}(\lambda_{31} + \lambda'_{31} +  2\lambda_3)v^2, 
\quad \Lambda_{\phi_2}= \frac{1}{2}(\lambda_{23} + \lambda'_{23} +2\lambda_2 )v^2  
\qquad \qquad 
\eea

\item \textbf{The charged inert fields}\\
The pair of inert charged gauge eigenstates, $\phi^\pm_{1}, \phi^\pm_{2}$, which are rotated by
\be 
R_{\theta_c}= 
\left( \begin{array}{cc}
\cos \theta_c & \sin \theta_c \\
-\sin \theta_c & \cos \theta_c\\
\end{array} \right), \qquad \mbox{with} \quad \tan 2\theta_c = \frac{2\mu^2_{12}}{\mu^2_1 - \Lambda'_{\phi_1} - \mu^2_2 + \Lambda'_{\phi_2}} \nonumber
\ee
into the mass eigenstates, $H^\pm_1, H^\pm_2$, have squared masses:
\bea
&& m^2_{H^\pm_1}=  (-\mu^2_1 + \Lambda'_{\phi_1})\cos^2\theta_c + (- \mu^2_2 + \Lambda'_{\phi_2}) \sin^2\theta_c -2\mu^2_{12} \sin\theta_c \cos\theta_c \nonumber\\
&& m^2_{H^\pm_2}= (-\mu^2_1 + \Lambda'_{\phi_1})\sin^2\theta_c + (- \mu^2_2 + \Lambda'_{\phi_2}) \cos^2\theta_c + 2\mu^2_{12} \sin\theta_c \cos\theta_c \nonumber\\
&& \quad  \mbox{where} \quad \Lambda'_{\phi_1}= \frac{1}{2}(\lambda_{31})v^2  , 
\quad \Lambda'_{\phi_2}= \frac{1}{2}(\lambda_{23} )v^2  
\qquad \qquad \qquad \qquad \qquad \qquad \qquad \quad 
\eea

\item \textbf{The CP-odd neutral inert fields}\\
The pair of inert pseudo-scalar gauge eigenstates, $A^0_{1}, A^0_{2}$, which are rotated by
\be 
R_{\theta_a}= 
\left( \begin{array}{cc}
\cos \theta_a & \sin \theta_a \\
-\sin \theta_a & \cos \theta_a\\
\end{array} \right), \qquad \mbox{with} \quad \tan 2\theta_a = \frac{2\mu^2_{12}}{\mu^2_1 - \Lambda''_{\phi_1} - \mu^2_2 + \Lambda''_{\phi_2}}\nonumber
\ee
into the mass eigenstates, $A_1, A_2$, have squared masses: 
\bea
&& m^2_{A_1}= (-\mu^2_1 + \Lambda''_{\phi_1})\cos^2\theta_a + (- \mu^2_2 + \Lambda''_{\phi_2}) \sin^2\theta_a -2\mu^2_{12} \sin\theta_a \cos\theta_a \nonumber\\
&& m^2_{A_2}= (-\mu^2_1 + \Lambda''_{\phi_1})\sin^2\theta_a + (- \mu^2_2 + \Lambda''_{\phi_2}) \cos^2\theta_a + 2\mu^2_{12} \sin\theta_a \cos\theta_a \nonumber\\
&& \quad \mbox{where} \quad \Lambda''_{\phi_1}= \frac{1}{2}(\lambda_{31} + \lambda'_{31} - 2\lambda_3)v^2 , 
\quad \Lambda''_{\phi_2}= \frac{1}{2}(\lambda_{23} + \lambda'_{23} -2\lambda_2 )v^2   
\qquad \qquad 
\eea

\end{itemize}

We will refer to $(H_1,A_1,$ $H^\pm_1)$ as the fields from the first generation and to
$(H_2,A_2,H^\pm_2)$ as the fields from the second generation.
Each of the four neutral particles could, in principle, be the DM candidate, provided it is lighter than the other neutral  states. In what follows, without loss of generality, 
we assume the CP-even\footnote{For the CP-even particle to be the DM candidate rather than the CP-odd particle, it is required that $m_{H_1} < m_{A_1}$, which leads to $\lambda_2,\lambda_3<0$. If instead $A_1$ is assumed to be the DM candidate, $\lambda_2,\lambda_3>0$. Hence, the results of our analysis are also applicable to the $A_1$ DM case by changing the sign of $\lambda_2$ and $\lambda_3$.} neutral particle $H_1$ 
from the first generation to be lighter than all other inert particles, that is:
\begin{equation}
m_{H_1} < m_{H_2}, m_{A_{1,2}},m_{H^\pm_{1,2}}.
\end{equation}
(Note that this choice is arbitrary: if the CP-even particle from the second generation, $H_2$, where to be assumed lighter than the other inert states, then $H_2$ will play the role of the DM candidate.)

Assuming the CP-even neutral inert particles are lighter than the CP-odd and charged inert particles puts the following constraints on the parameters:
\be 
2\lambda_2 , 2\lambda_3 < \lambda'_{23}, \lambda'_{31} < 0. 
\ee
In our DM analysis, we consider cases where the mass alignment is changed, but where $H_1$ is always the lightest inert state and hence is the DM particle.
In the remainder of the paper the notations $H_1$ and DM particle
will be used interchangeably.

%\begin{figure}[ht!]
%\centering
%\includegraphics[scale=0.8]{Masses-3HDM.pdf} 
%\caption{Schematic mass-squared spectrum of the $Z_2$ symmetric I(2+1)HDM, where 
%$\Sigma= 4\mu^4_{12} + (\mu^2_1-\Lambda_{\phi_1} -\mu^2_2 +\Lambda_{\phi_2})^2 $, 
%$\Sigma' = 4\mu^4_{12} + (\mu^2_1-\Lambda'_{\phi_1} -\mu^2_2 +\Lambda'_{\phi_2})^2 $ and 
%$\Sigma''=4\mu^4_{12} + (\mu^2_1-\Lambda''_{\phi_1} -\mu^2_2 +\Lambda''_{\phi_2})^2 $. As discussed in the text, other mass orderings are considered in our DM analysis, while $H_1$ is always the lightest state.}
%\label{Masses-fig}
%\end{figure}

\subsection{Simplified couplings in the I(2+1)HDM}

Due to the large number of free parameters in the I(2+1)HDM which makes it impractical to analyse the model in the general case, we focus on a simplified case where parameters related to the first inert doublet are $k$ times the parameters related to the second doublet
\be 
\label{lambda-assumption} 
\mu^2_1 = k \mu^2_2, \quad \lambda_3 = k \lambda_2, \quad \lambda_{31} = k \lambda_{23}, \quad \lambda_{31}' = k \lambda_{23}', 
\ee
resulting in
\be 
\Lambda_{\phi_1} = k \Lambda_{\phi_2}, \quad \Lambda'_{\phi_1} = k\Lambda'_{\phi_2}, \quad \Lambda''_{\phi_1} = k \Lambda''_{\phi_2},
\ee
without introducing any new symmetry to the potential. The motivation for this simplified scenario is that in the $k=0$ limit the model reduces to the well-known I(1+1)HDM. We assume no specific relation among the other parameters of the potential. It is important to note that the remaining quartic parameters do not influencethe  discussed DM and LHC phenomenology of the model and thus their values have been fixed in agreement with  the theoretical constraints discussed in the original paper \cite{Keus:2014jha} and compliant with the results on unitarity obtained in \cite{Moretti:2015cwa}.

\subsubsection*{The $k=1$ case}
In this paper we focus on the $k=1$ case in the heavy DM mass region\footnote{Other scenarios with $k \neq 1$ are studied in \cite{Keus:2014jha}.}. The mass spectrum in this case is simplified to:
\bea 
&& m_{H_1}^2 = - \mu_2^2 + \Lambda_{\phi_2} - \mu_{12}^2, \qquad m_{H_2}^2 =  m_{H_1}^2 + 2\mu_{12}^2,\\
&& m_{H^\pm_1}^2 = - \mu_2^2 + \Lambda'_{\phi_2} - \mu_{12}^2, \qquad m_{H^\pm_2}^2 = m_{H^\pm_1}^2 + 2\mu_{12}^2, \nonumber\\
&& m_{A_1}^2 = - \mu_2^2 + \Lambda''_{\phi_2} - \mu_{12}^2 , \qquad m_{A_2}^2  = m_{A_1}^2 + 2\mu_{12}^2. \nonumber
\eea

The quartic couplings in the potential can  be written in terms of the masses of the physical particles as:
\bea
&& \lambda_{23}' = \frac{1}{v^2} (m_{H_1}^2 +m_{A_1}^2 -2m_{H^\pm_1}^2), \nonumber\\
&& \lambda_2 = \frac{1}{2v^2} (m_{H_1}^2 - m_{A_1}^2), \\
&& \lambda_{23} = g_{H_1 H_1 h} - \frac{2}{v^2}(m_{H_1}^2 - m_{H^\pm_1}^2), \nonumber
\eea
where $g_{H_1 H_1 h}= \lambda_{23}+\lambda_{23}'+2\lambda_2$ is the Higgs-DM coupling. The Feynman rules for this model are presented in Appendix \ref{Feyn-rules}.

%%%%%%%%%%%%%%%%%%%%%%%%%%%%%%%%%%%%%%%%%%%%%%%%%%%%%%%%%%%%%%%%%%%%%%%%%%%%%%
\section{Calculating the relic density in the I(2+1)HDM}\label{sec:annihilation}

The relic density of the WIMP (identified in our model as the lightest inert scalar 
$H_1$)
is calculated with the assumption that the WIMP was in thermal equilibrium with the SM particles after inflation. Once the rate of $$\mbox{DM DM}  \leftrightarrow  \mbox{SM  SM}  $$ reactions becomes smaller than the Hubble expansion rate of the Universe, the WIMP freezes out, i.e., drops out of the thermal equilibrium. After freeze-out the co-moving WIMP density remains essentially constant with the current value estimated by the Planck experiment to be the one
already given in Eq. (\ref{relic}).

As mentioned, in the I(2+1)HDM one of the neutral inert (pseudo-)scalar particles play the role of the DM. The relic density of a
(pseudo-) scalar DM candidate, $S$, after freeze-out is given by the solution to  the Boltzmann equation:
\be 
\frac{d n_S}{dt} = - 3 H n_S - \langle \sigma_{eff} v \rangle (n_S^2 - n^{eq \; 2}_{S}), 
\qquad S=H_1, H_2, A_1,A_2,
\ee
where the thermally averaged effective (co)annihilation cross-section contains all relevant scattering  processes of any $S_i S_j$ pair into SM particles:
\be 
\langle \sigma_{eff} v \rangle = \sum_{ij} \langle \sigma_{ij} v_{ij} \rangle \frac{n^{eq}_i}{n^{eq}_S} \frac{n^{eq}_j}{n^{eq}_S},
\ee
where
\be 
\frac{n^{eq}_i}{n^{eq}_S} \sim \exp({-\frac{m_i - m_S}{T}}).
\ee
Therefore, only processes in which the mass splitting between a state $S_i$ and the lightest $Z_2$-odd particle $S$ ($H_1$ in our case) are comparable to the thermal bath temperature $T$ provide a sizeable contribution to this sum.

In the I(2+1)HDM, the presence of additional inert particles has important consequences in the heavy mass regime. For lighter masses the most important channel for the annihilation of DM  particles is the Higgs-mediated process 
$$H_1 H_1 \to f \bar{f}$$
(see Fig. \ref{diag1}a), as studied in \cite{Keus:2014jha}. However, coannhilation with $H_2,A_1$ and $A_2$ may change the results significantly (see Fig. \ref{diag1}b). 

For heavier masses the diagrams including one or two virtual gauge bosons, shown in Fig. \ref{diag2} also contribute to the total annihilation cross-section. Finally, co-annihilation plays an important role in scenarios with multiple particles which are close in mass. This scenario is realised in the I(2+1)HDM for the heavy DM mass region. Particles up to 20\% heavier than the DM candidate may influence the DM relic density. Therefore, the co-annihilation diagrams should be included in calculating the effective annihilation cross-section. These diagrams are presented in Figs. \ref{diag3} and \ref{diag4} -- representing pure gauge channels and coannhilation channels involving the SM-like Higgs particle, respectively.

\subsection{Relevant co-annihilation scenarios}
%\label{annihilation-scenarios}

In the I(2+1)HDM, the strength and importance of coannhilation processes depend on the mass splittings between the inert particles. We define $\delta_A$ and $\delta_C$ as the splitting between $H_1$ and the pseudoscalar and charged state from the first generation, respectively,
\be 
\delta_A = m_{A_1} -m_{H_1} , \quad \delta_C = m_{H^\pm_1} - m_{H_1}.
\ee
$\delta_{A,C}$ are related to the quartic couplings in the potential, which are constrained by the perturbativity (and unitarity) conditions, i.e., the $\lambda_i$'s cannot be too large. As a result of this, all particles within one generation will have a similar mass. These masses, however, could have high values because of the (almost) unconstrained quadratic parameters $\mu_{2}^2$ and $\mu_{12}^2$:
\bea
&&  m_{H_1}^2 = -\mu^2_2 - \mu^2_{12} +\frac{v^2}{2} g_{H_1 H_1 h}. 
\eea

It is important to stress that, even if bounds on $\lambda_i$ were relaxed leading to larger values of $\delta_{A,C}$, there exist very stringent limits from relic density analysis. Coannihilation must occur at least between $H_1, A_1$ and $H^\pm$ to achieve DM relic density in agreement with the current experimental measurements. This is a pattern followed by all general heavy scalar DM models. In the absence of these co-annihilation channels, the maximum relic density that can be achieved through $H_1 H_1 \to \textrm{ SM SM}$ (even when $H_i H_j \to \textrm{ SM SM}$ is allowed) is of order $10^{-3}$ which is well below the observed value.

The other important mass splitting, $\Delta$, is defined as the mass difference between $H_1$ and the other CP-even state $H_2$ (``splitting between doublets''):
\be 
\Delta = m_{H_2} - m_{H_1}
\ee
$\Delta$ is related to the quadratic parameter $\mu_{12}^2$ through
\be
\mu_{12}^2 = \frac{1}{2}(m_{H_2}^2 - m_{H_1}^2) = \frac{1}{2} \left(\Delta^2+2 m_{H_1} \Delta\right).
\ee

Note that $\mu_{12}^2$ is not limited by any theoretical constraints -- similar to $\mu_{2}^2$ -- and therefore $\Delta$ can in principle be very large. So, unless $\Delta$ is forced to be small by limits put on $\mu_{12}^2$, one should also consider a case where the second doublet is decoupled from the first, leading to a scenario which was not listed in \cite{Keus:2014jha}. Therefore, in the very heavy mass region one can consider:
\begin{itemize}
\item \textbf{Case G}: with small $\delta_A, \delta_C, \Delta$,
where all inert particles are close in mass and co-annihilate with each other.

\item \textbf{Case H}: with small $\delta_A, \delta_C$ and large $\Delta$,
where the second generation is effectively decoupled from the first generation and does not influence relic density calculations. In this case, the relevant diagrams are the (co)annihilation channels between fields from the lighter generation only, $H_1,A_1,H_1^\pm$. 
\end{itemize}

Tab. \ref{scenarios} summarises the two possible scenarios for relic density studies.

\begin{table}[h]
\centering
\begin{tabular}{| c | c | m{8cm} |} \hline
 & Small $\delta_{A},\delta_{C}$  & Large $\delta_{A},\delta_{C}$   \\[0.1cm] \hline 
 Small $\Delta$ & Case G is realised  & Coannhilation between $H_1,H_2$ is not efficient enough and DM density is below experimental bounds. \\[0.2cm] \hline
Large $\Delta$ & Case H is realised & There are no co-annihilatin channels open and gauge annihlation reduces DM relic density effectively below the experimental bounds. \\[0.2cm] \hline 
\end{tabular}
\caption{Valid regions of the parameter space schematically shown in terms of $\delta_{A},\delta_{C}$ and $\Delta$ in the heavy mass regime of I(2+1)HDM.}
\label{scenarios}
\end{table}

\subsection{The gauge limit}
To illustrate the difference between cases G and H, let us first consider the gauge limit in both scenarios, which is the limit where all quartic couplings $\lambda_i$ are set to zero. Therefore, all scalar self-couplings, including the DM-Higgs coupling, are removed in this limit. As a result $\delta_{A,C} = 0$, leading to degenerate $H_1,A_1,H^\pm_1$ states. Note that this limit is excluded by results of direct detection experiments, nevertheless, it is an interesting limit to study as it represents the main difference between cases G and H. In this limit $H_1$ annihilates solely through the gauge annihilation channels presented in Fig. \ref{diag3}.

Non-zero $\lambda_i$ will lift this degeneracy and, at the same time, reduce the effective annihilation cross-section for a given mass. Therefore, the gauge limit corresponds to the minimum value of $m_{H_1}$, for which it is possible to obtain a proper relic density for any value of the Higgs-DM coupling. 

\begin{figure}[h!]
\centering
\includegraphics[scale=1]{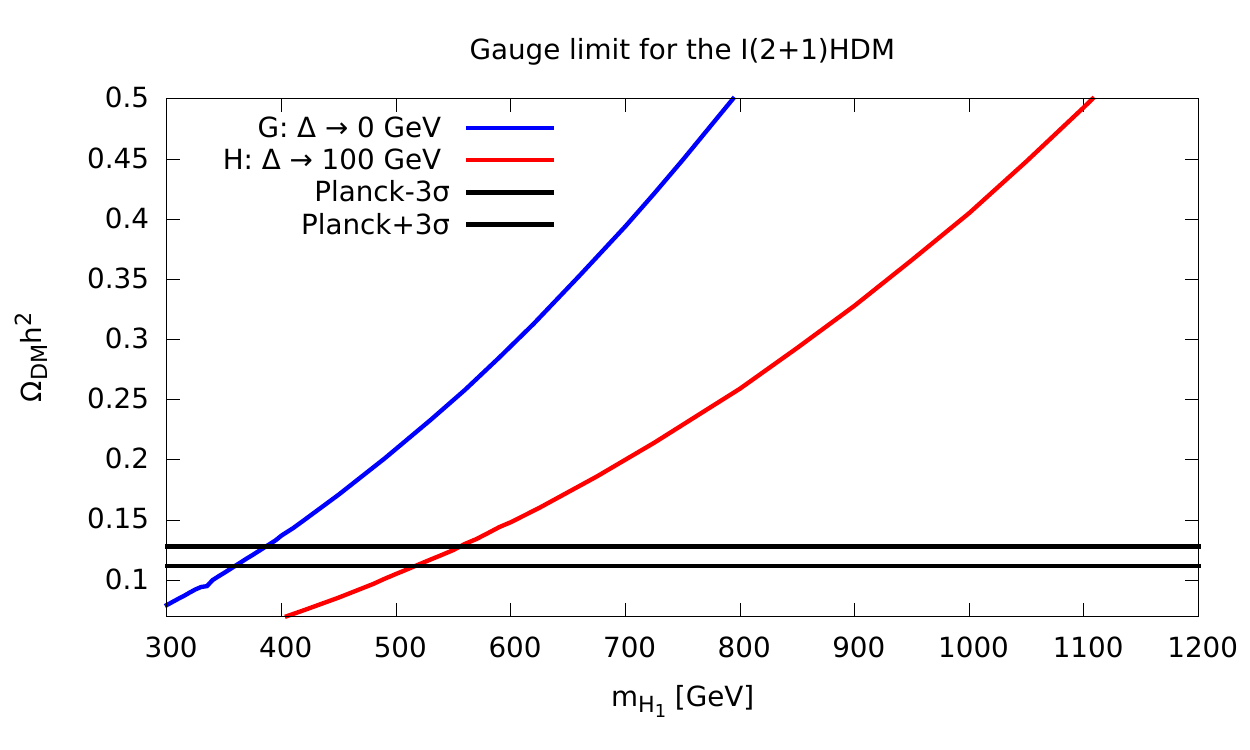}
\caption{The gauge limit in case G (blue curve) where all particles have degenerate mass (in the $\delta_{A,C}, \Delta \to 0$) and case H (red curve) with large $\Delta$ where the second generation is decoupled from the first generation and all particles in the first generation are degenerate in mass (in the $\delta_{A,C} \to 0$ limit).}
\label{gauge-limit}
\end{figure}

These results are presented in Fig. \ref{gauge-limit} for the two scenarios: case G where all particles have degenerate mass (in the $\delta_{A,C}, \Delta \to 0 \Leftrightarrow$ $\lambda_i \to 0, \mu_{12}^2 \to 0$ limit) and case H with large $\Delta$ where the second generation is decoupled from the first generation and all particles in the first generation are degenerate in mass (in the $\delta_{A,C} \to 0 \Leftrightarrow \lambda_i \to 0$ limit). 

It is clear that, for a given mass of $m_{H_1}$ the destructive interference between an increased number of coannhilation diagrams in case G leads to a reduced cross-section, i.e. larger DM relic density with respect to case H. The important consequence  of this interference is that for case G it is possible to obtain proper relic density for smaller masses of DM candidate in comparison to case H. Note also that case H behaves like the I(1+1)HDM (the Inert Doublet Model) in this limit. This similarity in behaviour will be repeated as we will show in the following sections.

%Limits for masses: 360 GeV for case G, 519 GeV for case H.

\subsection{The benchmark points}
With non-zero scalar couplings and mass splitting more diagrams contribute to the co-annihilation of DM -- all diagrams shown in Figs. \ref{diag3} and \ref{diag4} contribute to the total annihilation cross-section. Here we present two benchmarks points, two sets of parameters, for which we have studied the DM relic density:
% ... plot for G, plot for H, plot for GandH for one mass, a dolphin plot

\begin{itemize}
\item \textbf{For case G with $\delta_A = \delta_C = 1$ GeV and $\Delta = 1$ GeV}\\
Here all inert particles have similar masses and therefore can co-annihilate with each other. The degeneracy between charged and ``pseudo-scalar" particles is allowed and doesn't lead  to any unacceptable results. The important degeneracy which must be avoided is $H_1$-$A_1$ degeneracy leading to the scattering through the $Z$ boson which is tightly constrained by direct detection experiments and puts a lower limit on $\delta_A$.

\item \textbf{For case H with $\delta_A = \delta_C = 1$ GeV and $\Delta = 100$ GeV}\\
Here the second generation of inert scalars is significantly heavier than the first one. Within each generation, however, particles are almost degenerate. 
\end{itemize}

\begin{figure}[h!]
\centering
\includegraphics[scale=0.8]{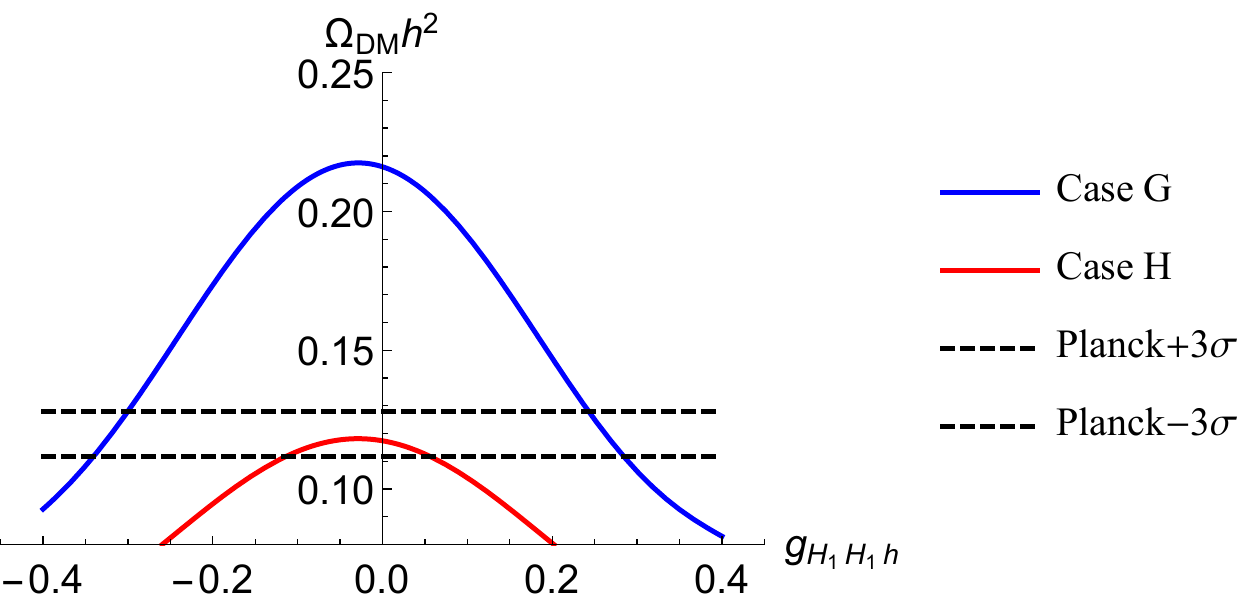}
\caption{Relic density plots in case G ($\Delta = 1$ GeV, blue line) and H ($\Delta = 100$ GeV, red line) for $m_{H_1} = 550$ GeV. The dashed horizontal lines show the 3$\sigma$ relic density limits from Planck in Eq.
    (\ref{relic}).}
\label{relic550}
\end{figure}

Note that there are certain differences between cases G and H.
In case H, the heavier generation of inert particles is decoupled from the first generation particles and does not influence the relic density calculations. The model in this case resembles the I(1+1)HDM.
Furthermore, in case G the Higgs-DM couplings which result in a relic density in agreement with experiment are larger in comparison to case H for the same DM mass. This difference is explicit in Fig. \ref{relic550}.

\begin{figure}[h!]
\centering
\includegraphics[scale=0.6]{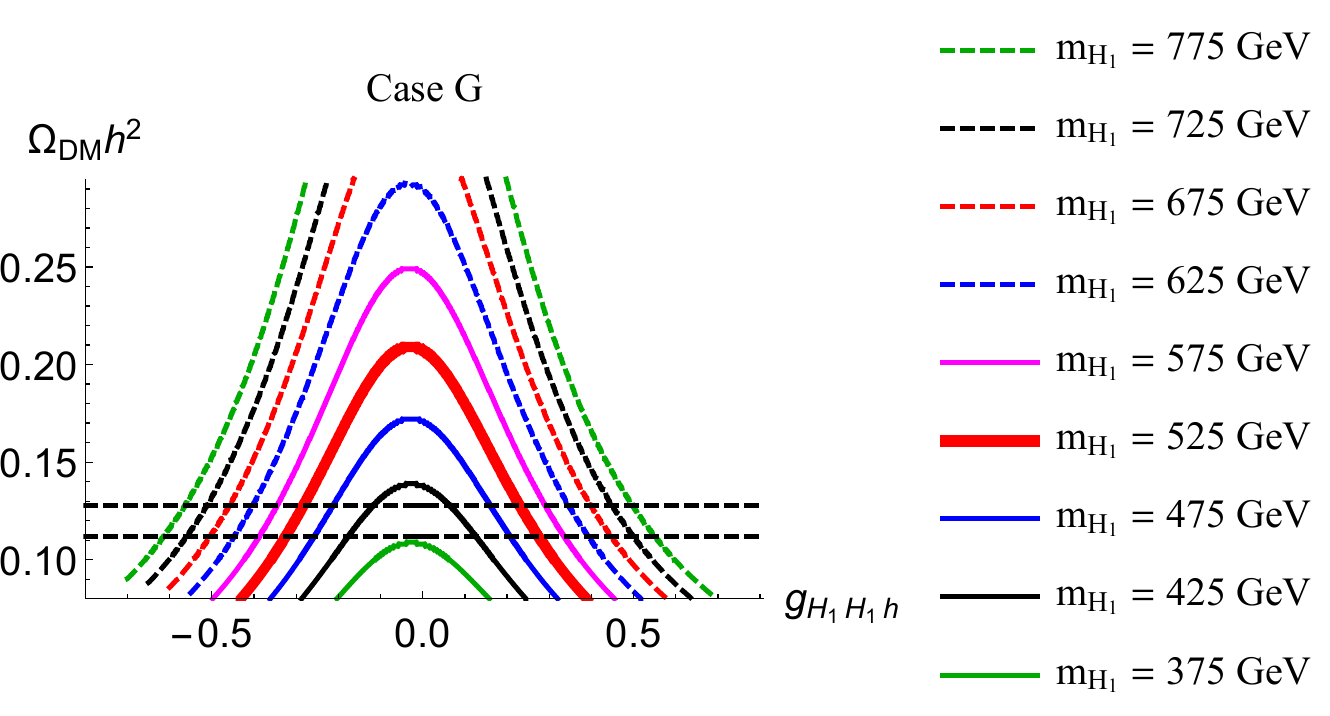}
\includegraphics[scale=0.6]{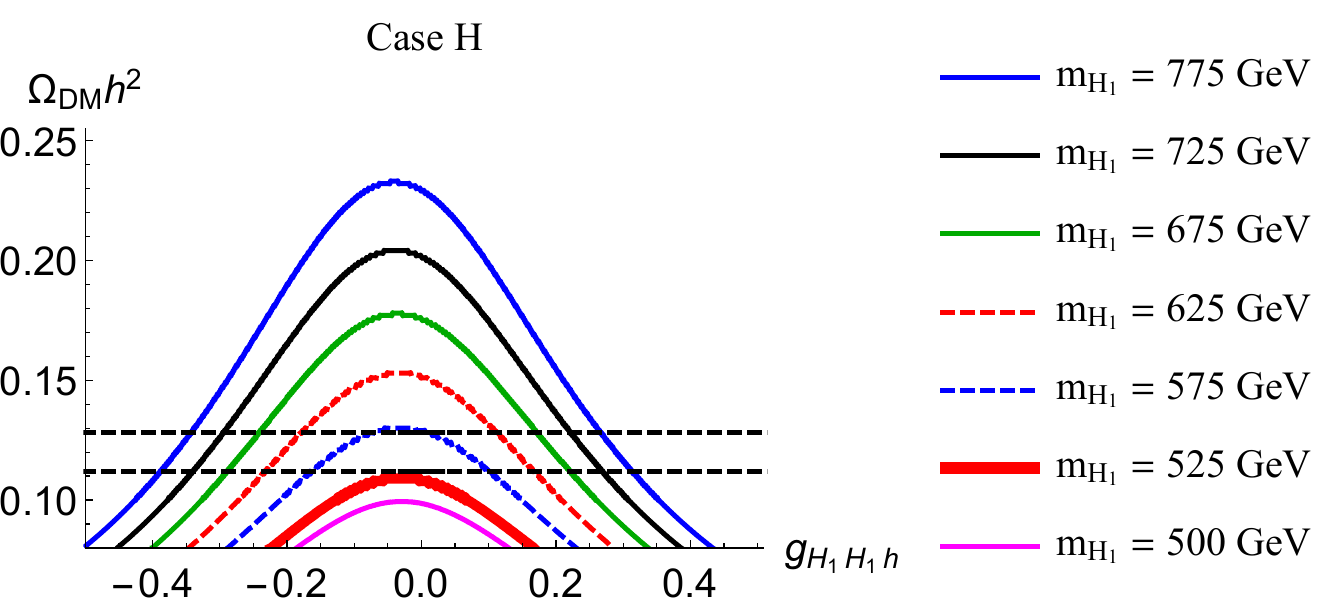}
\caption{Relic density plots for case G (left) with $\Delta =1$ GeV, $\delta =1$ GeV and case H (right) with $\Delta =100$ GeV, $\delta =1$ GeV. Note the solid red line in the case H which represents the minimum DM mass $m_{H_1}=525 \gev$ which just touching the lower relic density limit. The relic density plot for the same DM mass has been highlighted in case G (red solid line) which is well within the acceptable relic density limits. The dashed horizontal lines show the 3$\sigma$ relic density limits from Planck in Eq. (\ref{relic}).} 
\label{relicHG}
\end{figure}

This leads to the fact that for case G we can obtain viable relic density values for $m_{DM}$ much smaller than in case H (or the I(1+1)HDM) in which the minimal value of $m_{DM}$ resulting in DM relic density in agreement with Planck limits is $m_{H_1} \approx 525-535$ GeV. In case G (with $\Delta = 1$ GeV), however, the DM mass can be as low as $\sim 375$ GeV. This result is shown in Fig. \ref{relicHG} which represents relic density plots for cases G (left) and  H (right). Note that in case G the minimum $m_{H_1}$ which touches the lowest acceptable relic density limit (the green solid line) is $375$ GeV (for a given $\Delta$ of 1 GeV), whereas in case H this minimum value is 525 GeV (the solid red line). 

%\begin{figure}[h!]
%\centering
%\includegraphics[scale=0.6]{heavy3hdmH.pdf}
%\includegraphics[scale=0.6]{heavy3hdmG.pdf}
%\caption{Left: Case H ($\Delta = 100$ GeV). Right: Case G ($\Delta = 1.5$ GeV)}
%\label{relicHG}
%\end{figure}

\begin{figure}[h!]
\centering
\includegraphics[scale=0.7]{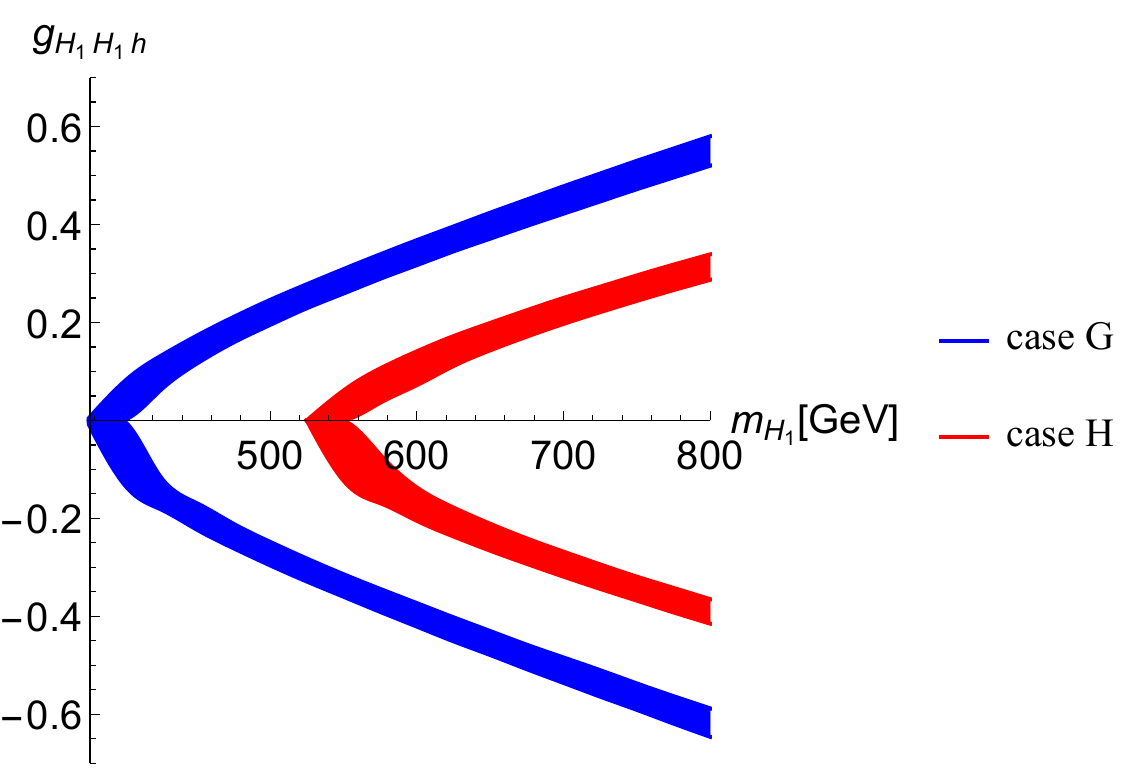}
\caption{Relic density bands in agreement with Planck measurements in case G (for an exemplary $\Delta = 1$ GeV, $\delta = 1$ GeV) in red and  case H (for an exemplary $\Delta = 100$ GeV, $\delta = 1$ GeV) in red. Note that for the a given DM mass, the Higgs-DM coupling in case G is much larger than in case H.}
\label{relicHG2}
\end{figure}

Fig. \ref{relicHG2} is meant to represent the same benchmark points as in Fig. \ref{relicHG}, in the $m_{H_1}$-$g_{H_1 H_1 h}$ plane. The bands correspond to proper relic density in agreement with Planck measurements in case G (for an exemplary $\Delta = 1$ GeV, $\delta = 1$ GeV) in red and  case H (for an exemplary $\Delta = 100$ GeV, $\delta = 1$ GeV) in red. Note that, for the a given DM mass (and same $\delta$), the Higgs-DM coupling in case G is much larger than in case H. 

%This is the lower limit for $m_{H_1}$ for those $\delta_{A,C},\Delta$. Clearly, there is a new region with respect to the IDM.

%\subsection{Variation of parameters: different values of DM mass}
%
%We can go up with masses easily up to a few TeV, probably higher (up to 50 TeV) . However, there are two consequences: first, the particles are getting heavier and "more decoupled" from the SM, and therefore probably less and less possible to detect. Second, heavier DM masses mean bigger coupling needed to obtain proper relic density value -- for example, 20 TeV particle may require $g_{h H_1 H_1} \approx 10$.  However, sensitivities of all experiments drop significantly with the rising DM mass, therefore there shouldn't be problems with agreement with direct- and indirect-detection experiments.

\section{DM in the I(2+1)HDM: direct detection }\label{sec:DM}

\subsection{Changes in $\Delta, ~\delta_{A}, ~\delta_{C}$}

%For the two viable scenarios G (small $\delta_{A,C},\Delta$) and H (small $\delta_{A,C}$ and big $\Delta$), where \textit{small} and \textit{big} refer to \textit{with} and \textit{without} co-annihilation with respective particles, we allow for the physical parameters to vary in the following regions:

For the two viable scenarios G and H  we allow the physical parameters to vary in the following regions:
\be 
100 \textrm{ keV} \lesssim \delta_A, \delta_C \lesssim 15\gev.
\label{delta-bounds}
\ee

Since $\delta_{A}$ and $\delta_{C}$ are related to the quartic parameters $\lambda_i$'s, they are constrained from unitarity bounds and are required to be small. However, regardless of any limits on $\lambda_i$ from unitarity, relic density studies show that both $\delta_A$ and $\delta_C$ must be relatively small to allow for co-annihilation between particles which is crucial in the heavy DM mass region.  The upper limit follows the following rough rule: co-annihilation effects take place when the mass difference between co-annihilating particles is of the order of 20\% of their mass.

%The boundary is $\delta_{A,C} \lesssim 15$ GeV, preferred [=less fine-tuned parameter space] smaller values of $\delta_{A,C}$. 

The lower bound on $\delta_A$ comes from direct detection experiments where a degeneracy between $H_1$ and $A_1$ leads to the scattering through the $Z$ boson which is tightly constrained. Further, the above limits in Eq.  
 (\ref{delta-bounds}) are in agreement with LEP searches for exotic particles. Finally, below $\delta_{A,C} \sim 0.1$ GeV there is no visible difference in the results.

As mentioned before, a large difference between $\delta_A$ and $\delta_C$ violates relic density limits. In the cases we study below, we have set $\delta_A = \delta_C = \delta$ for simplicity\footnote{Cases with $\delta_A \neq \delta_C$ do not lead to any new phenomenology and in fact the region of validity for $g_{H_1H_1h}$ decreases as we increase the difference between $\delta_A$ and $ \delta_C$.}. The mass splitting between the two generations, $\Delta$, is proportional to  and therefore unconstrained by unitarity. The maximum value for $\Delta$ is proportional to the arbitrary maximum value chosen for $\mu_{12}^2$. In general, we allow for $\Delta$ to vary in the following region
\be 
100 \textrm{ keV} \lesssim \Delta \lesssim 200 \gev.
\ee

For large $\Delta$ values, $\Delta \gtrsim 20-50$ GeV (depending on $m_{H_1}$, since the story-changing mass splitting is roughly 20\%$m_{H_1}$), co-annihilation effects between the two generations are not strong enough to compete with the standard (co)annhilation between $H_1,A_1,H^\pm_1$, in which case the I(2+1)HDM acts just like the I(1+1)HDM. The second generation is effectively decoupled from the first generation and does not influence DM phenomenology. Therefore, scenario H is realised for:
\be
\Delta \gtrsim 20 \gev \quad \Rightarrow \quad \mbox{scenario H}
\ee
%(Again - mass difference of roughly 20 \% of DM mass.)
The exact value of $\Delta$, when above $\sim 20-50$ GeV does not make any significant difference in the relic density calculations. 

For small values of $\Delta$, the co-annihilation effects between all particles are important. The smaller $\Delta$ is, the more relevant particles from the second generation are for DM studies: for $\Delta \approx 1.5 \gev$ the relative contribution to relic density calculation coming from particles from the lighter generation to the heavier generation is $70 \%$-$30 \%$. For $\Delta \approx 0.0001 \gev$ this relation is $50 \%$-$50\%$. The case G, is therefore realised when $\Delta$ varies in the following window
\be
0.0001 \gev \lesssim \Delta \lesssim 20 \gev \quad \Rightarrow \quad \mbox{scenario G}
\ee
The closer $\Delta$ gets to this upper limit, the weaker the coannnihilation effects and the more scenario H is realised.
Finally, notice that, for our studies, the Higgs-DM coupling, $g_{h H_1 H_1}$, is kept within the $|g_{h H_1 H_1}| < 1$ range.

\begin{figure}[h!]
\centering
\includegraphics[scale=0.6]{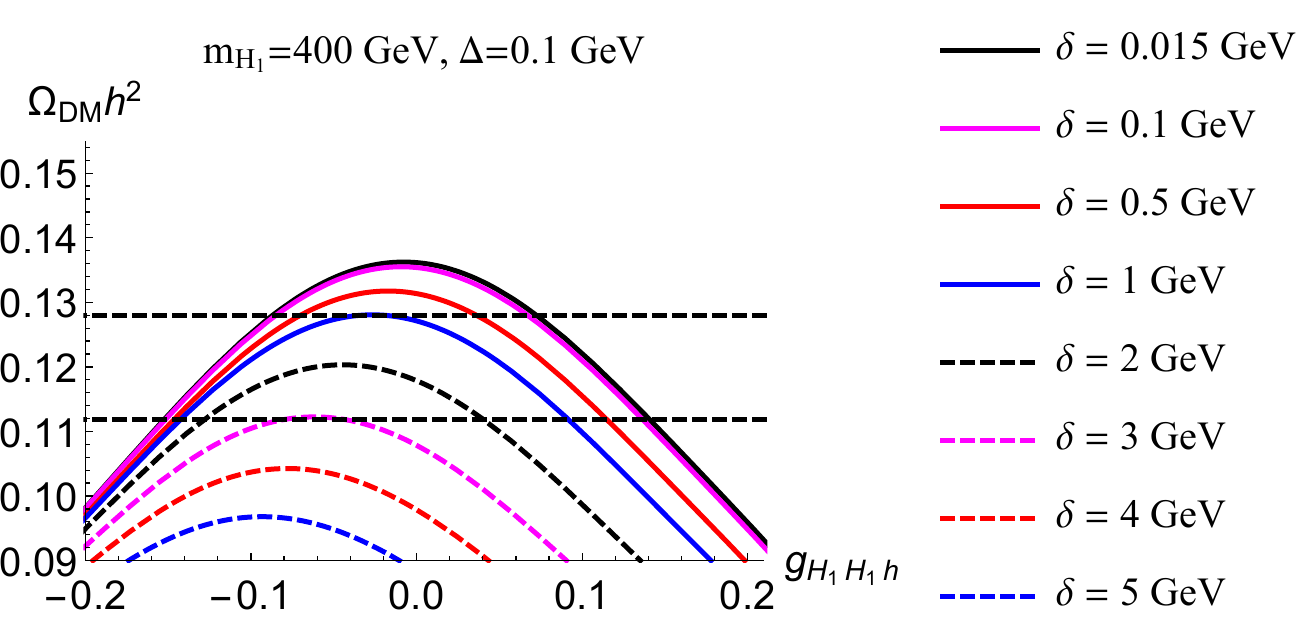}
\includegraphics[scale=0.6]{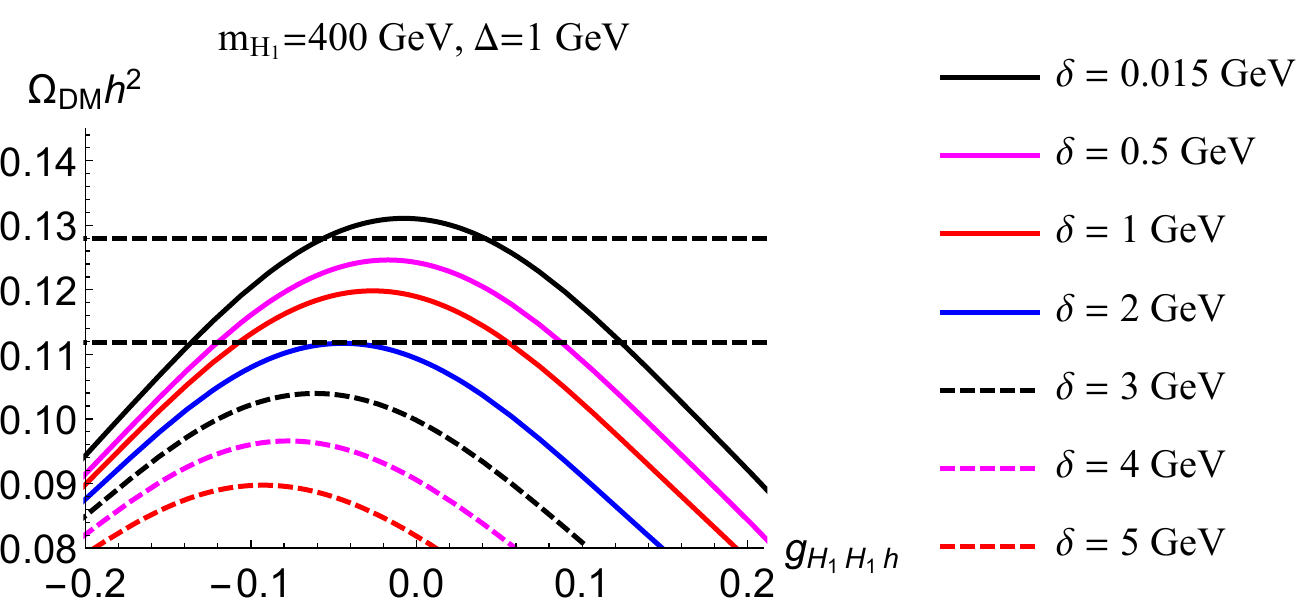}\\
\includegraphics[scale=0.6]{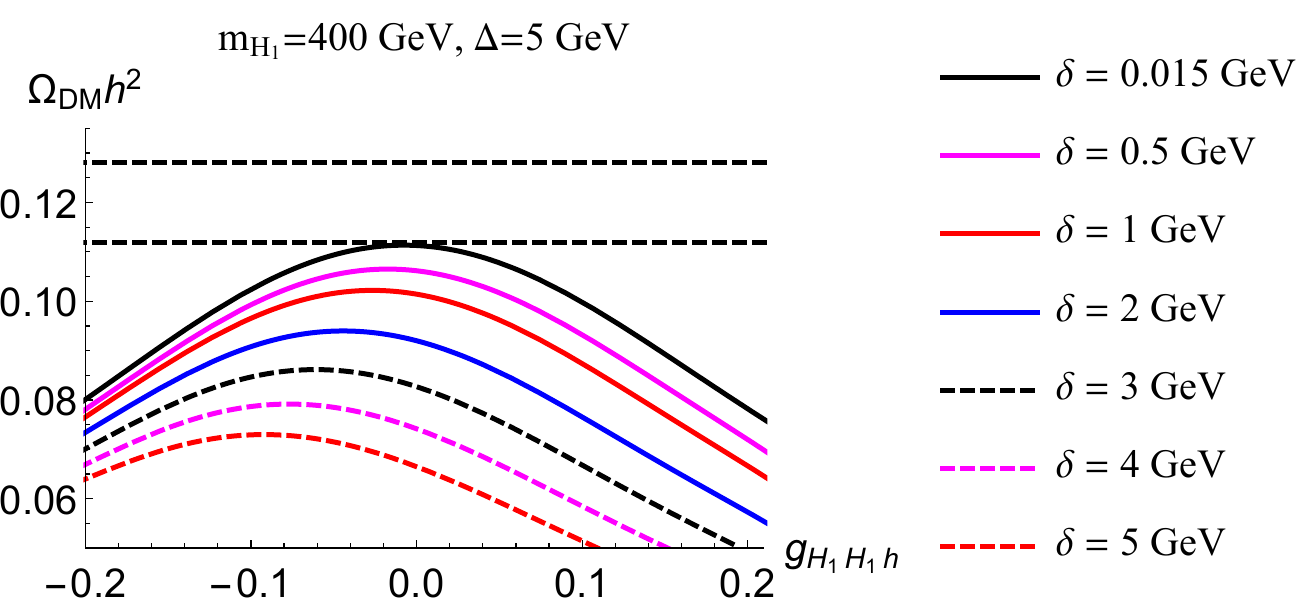}
\includegraphics[scale=0.6]{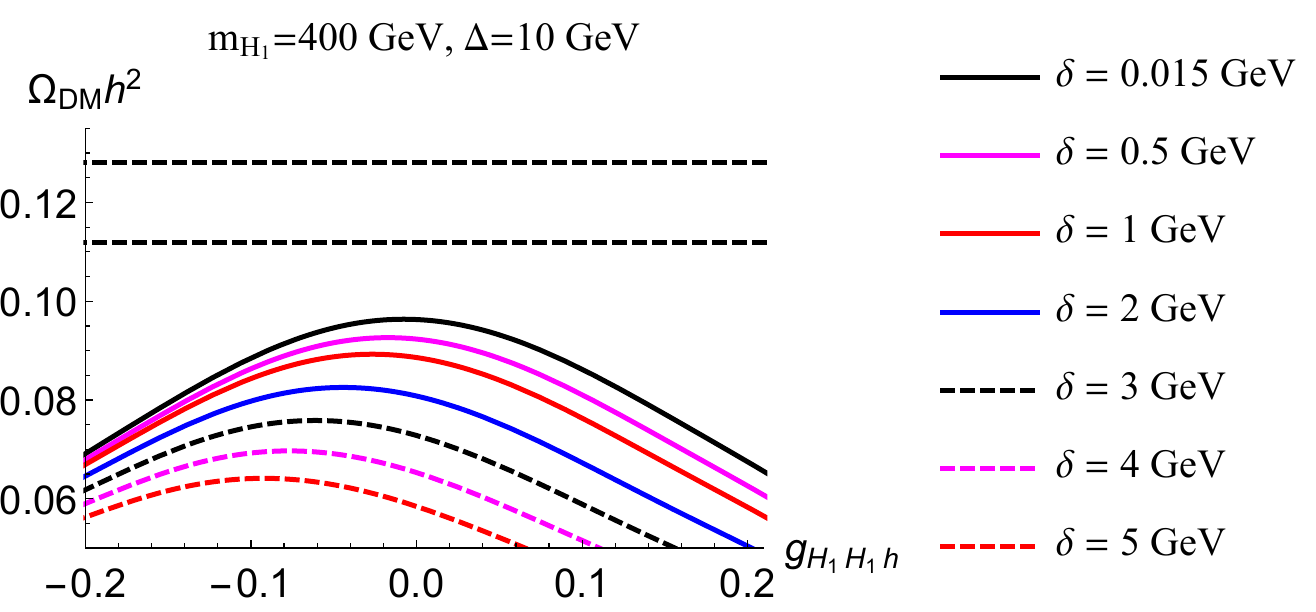}\\
\caption{The effect of changing $\Delta$ on the relic density. In all four plots $m_{DM}$ has been set to 400 GeV as the value of $\Delta$ changes, 0.1 GeV in the top left plot, 1 GeV in the top right plot, 5 GeV in the bottom left plot and 10 GeV in the bottom right plot. In each plot different colours represent different $\delta$s. For small values of $\Delta$ (0.1 GeV), the $H_1$-$H_2$ co-annihilation leads to viable relic density values even for large $\delta$ (i.e. the $H_1, A_1, H^\pm$ co-annihilation is absent). For large values of $\Delta$ (10 GeV) $H_1$-$H_2$ co-annihilation does not exist and even small values of $\delta$ cannot compensate this lack, thus, the relic density is below the acceptable limit.}
\label{relic-vs-g}
\end{figure}

Fig. \ref{relic-vs-g} illustrates the effect of changing $\Delta$ on the relic density. In all four plots $m_{DM}$ has been set to 400 GeV as the value of $\Delta$ changes, 0.1 GeV in the top left plot, 1 GeV in the top right plot, 5 GeV in the bottom left plot and 10 GeV in the bottom right plot. In each plot different colours represent different $\delta$s. For small values of $\Delta$ (0.1 GeV), the $H_1$-$H_2$ co-annihilation leads to viable relic density values even for large $\delta$ (i.e. the $H_1, A_1, H^\pm$ co-annihilation is absent). For large values of $\Delta$ (10 GeV) $H_1$-$H_2$ co-annihilation does not exist and even small values of $\delta$ cannot compensate this lack, thus, the relic density is below the acceptable limit.

\begin{figure}[h!]
\centering
\includegraphics[scale=0.6]{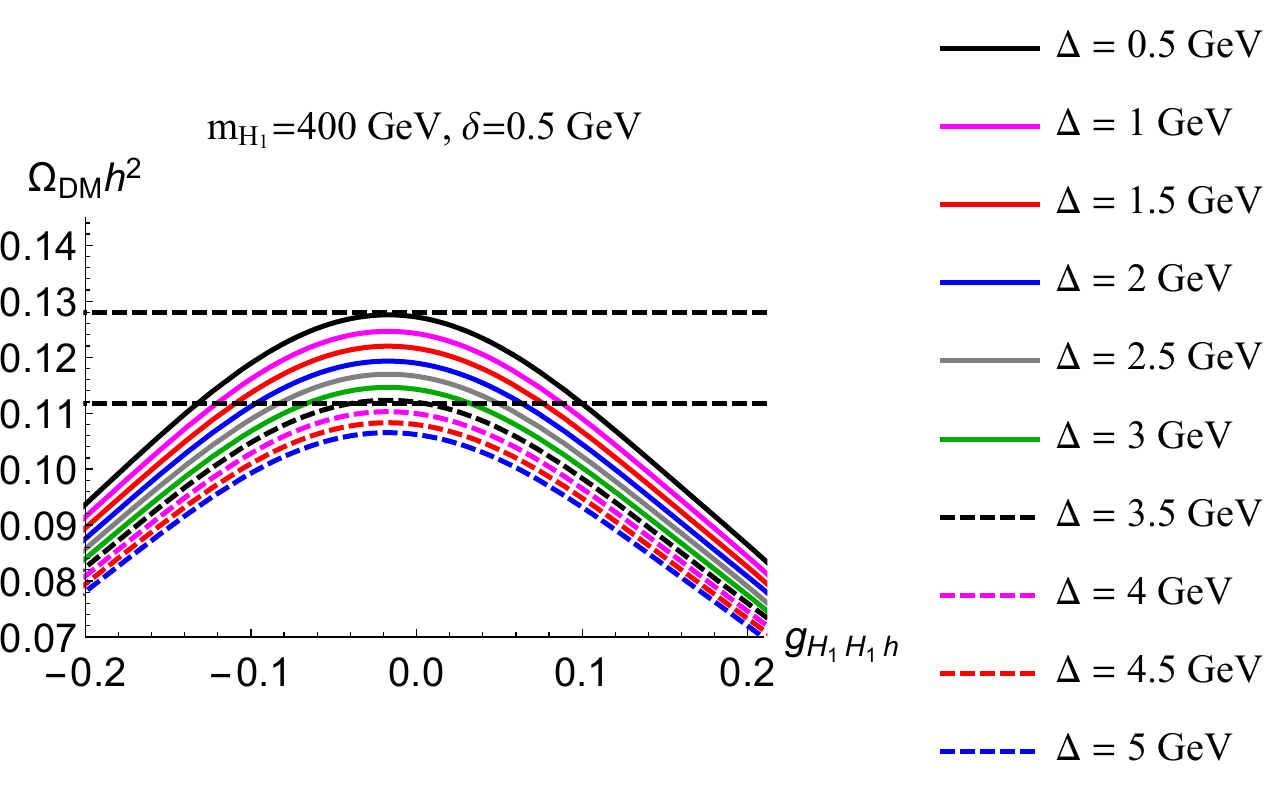}
\includegraphics[scale=0.6]{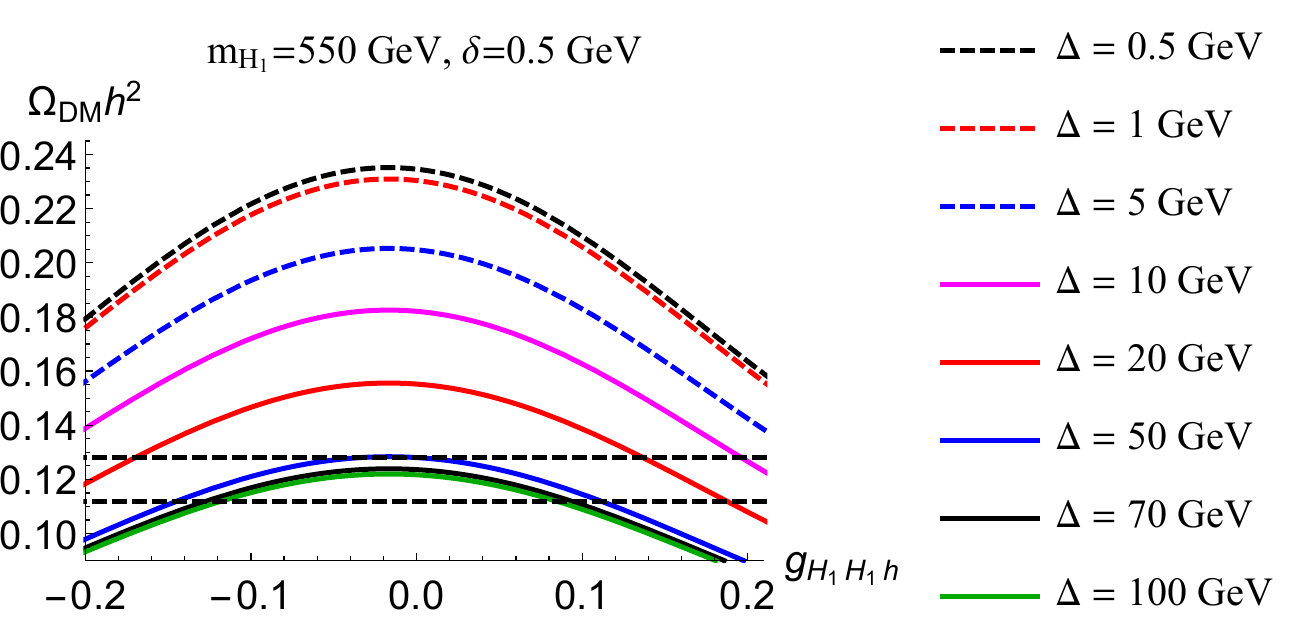}
\caption{Relic density plot for $\delta=0.5 \gev$ for $m_{DM}=400 \gev$ in the left plot and $m_{DM}=550 \gev$ in the right plot. In each plot, different colours correspond to changing $\Delta$s. Note that in the left plot only small values of $\Delta$ lead to viable relic density values, which is where case G is realised. In the right plot, small values of $\Delta$ correspond to case G and large values of $\Delta$ correspond to case H, and they all lead to acceptable relic density values. Note that for $\Delta \gtrsim 50$ GeV all curves correspond to the same value.}
\label{deltas}
\end{figure}

In Fig. \ref{deltas} the value of $\delta$ is set to $0.5 \gev$ for $m_{DM}=400 \gev$ in the left plot and $m_{DM}=550 \gev$ in the right plot. In each plot, different colours correspond to changing $\Delta$s. Note that in the left plot only small values of $\Delta$ lead to viable relic density values, which is where case G is realised. In the right plot, small values of $\Delta$ correspond to case G and large values of $\Delta$ correspond to case H, and they all lead to acceptable relic density values. Note that for $\Delta \gtrsim 50$ GeV all curves correspond to the same value.

\subsection{Changes in $m_{DM}$}
Here we describe several sub-regimes where the DM mass can vary with important characteristics.

\begin{itemize}
\item\
In the region 
\be 
m_{DM} \lesssim 360 \gev 
\ee
neither scenario H nor G results in viable relic density values\footnote{In extensions of the I(2+1)HDM with more inert doublets, this $m_{DM}$ limit could be lowered as more co-annihilation channels are present due to the extended number of inert particles.}. This lower limit can be reached in case G by very specific points in the parameter space: (a) when the mass splitting between all particles is tiny and all particles are almost degenerate in mass (up to $\mathcal{O} (100 \textrm{ keV})$ mass splitting to avoid direct detection limits), (b) when $g_{h H_1 H_1}$ is close to 0.

\item\
In the region 
\be 
360 \gev \lesssim m_{DM} \lesssim 525 \gev
\ee
only scenario G leads to acceptable values of DM relic density for specific values of $\Delta$ in the $0.0001 \gev \lesssim \Delta \lesssim 20 \gev$ and $|g_{ H_1H_1h}| \lesssim 0.3$ window. As a rule of thumb, smaller $\Delta$ allows for a wider viable region in the parameter space and the larger $m_{H_1}$ is the larger $|g_{h H_1 H_1}|$ must be. 

\item
In the region 
\be 
535 \gev \lesssim m_{DM} \lesssim 1.5-2 ~\mbox{TeV}
\ee
both scenarios, G and H, lead to viable values of DM relic density. The appropriate value of $g_{H_1H_1 h}$ coupling depends on the DM mass in each case. In scenario G, couplings are generally larger compared to scenario H and a larger DM mass requires larger values of the $g_{H_1H_1 h}$ coupling. 
%For example, limit $|g_{h H_1 H_1}| \lesssim 1$ corresponds to $M_H = 1.5$ TeV and $\Delta = 1.5 \gev$ (scenario G) and $M_H = 2$ TeV and $\Delta = 100 \gev$ (scenario H).
\end{itemize}

\begin{figure}[ht!]
\centering
\includegraphics[scale=0.8]{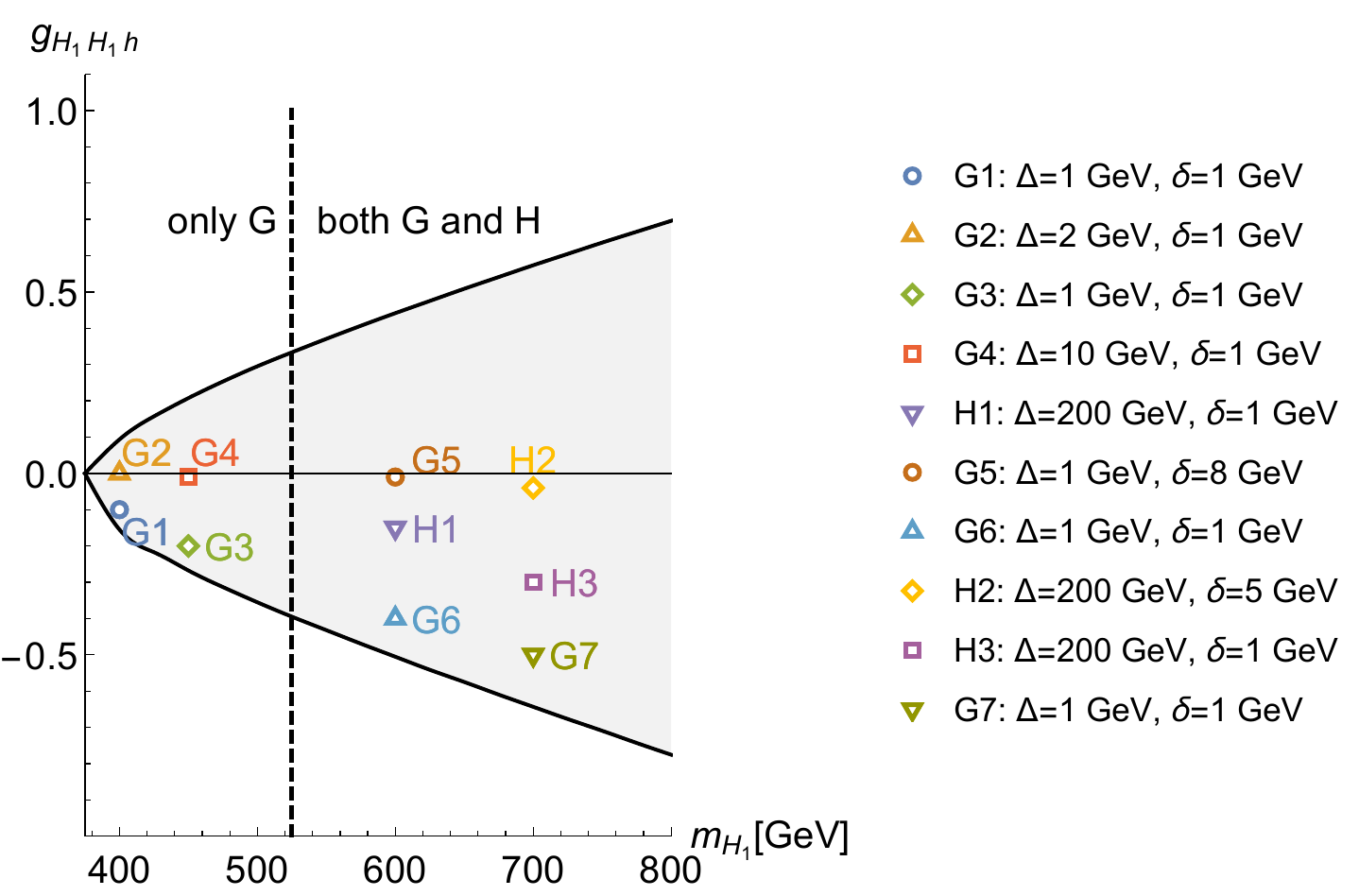}
\caption{Here G$i$ and H$i$ represent certain points in the parameter space corresponding to cases G and H, respectively. The shaded region is where the I(2+1)HDM has acceptable relic density results. To the left of the vertical dashed line case G is realised and to the right of it both cases G and H are realised. Generally the outermost parts of the shaded region are populated by G$i$ since they correspond to larger Higgs-DM couplings, and the innermost parts of the region correspond to case H. However, depending on the values of $\delta$ the Higgs-DM coupling changes and G$i$ points can appear close to the $g_{hH_1H_1}=0$ line as well, which is apparent in comparing the points G$5$, H$1$ and G$6$. Other points are shown on the plot for different values of $\Delta$ and $\delta$ for comparison. 
%(Dorota is producing a better version of this plot.)
} 
\label{flag}
\end{figure}

Fig. \ref{flag} is meant to summarise all that was said above in one plot with G$i$ and H$i$ representing certain points in the parameter space corresponding to cases G and H, respectively. The shaded region is where the I(2+1)HDM has acceptable relic density results. To the left of the vertical dashed line case G is realised and to the right of it both cases G and H are realised. Generally the outermost parts of the shaded region are populated by G$i$ since they correspond to larger Higgs-DM couplings whereas the innermost parts of the region correspond to case H. However, depending on the values of $\delta$ the Higgs-DM coupling changes and G$i$ points can appear close to the $g_{hH_1H_1}=0$ line as well, which is apparent in comparing the points G$5$, H$1$ and G$6$. Other points are shown on the plot for different values of $\Delta$ and $\delta$ for comparison.

\subsection{Direct detection}

Direct detection experiments, which are mostly designed to hunt for the standard EW-scale WIMP, are the most sensitive to DM masses of the order of 100 GeV. For heavier DM masses the sensitivity of these experiments drops significantly. The most recent LUX results set the limit of the DM-nucleon scattering cross-section to be $\sigma_{DM-N}\approx 10^{-8}$ pb for DM masses $\approx 500-1000$ GeV \cite{Akerib:2013tjd}. 

In the I(2+1)HDM, similar to the other scalar DM models, DM candidate can be detected through elastic scattering on nuclei through the exchange of a Higgs particle. Therefore, $\sigma_{DM-N}$ will depend on the value of DM-Higgs coupling, and the DM mass.

\begin{figure}[h!t]
\centering
\includegraphics[scale=1]{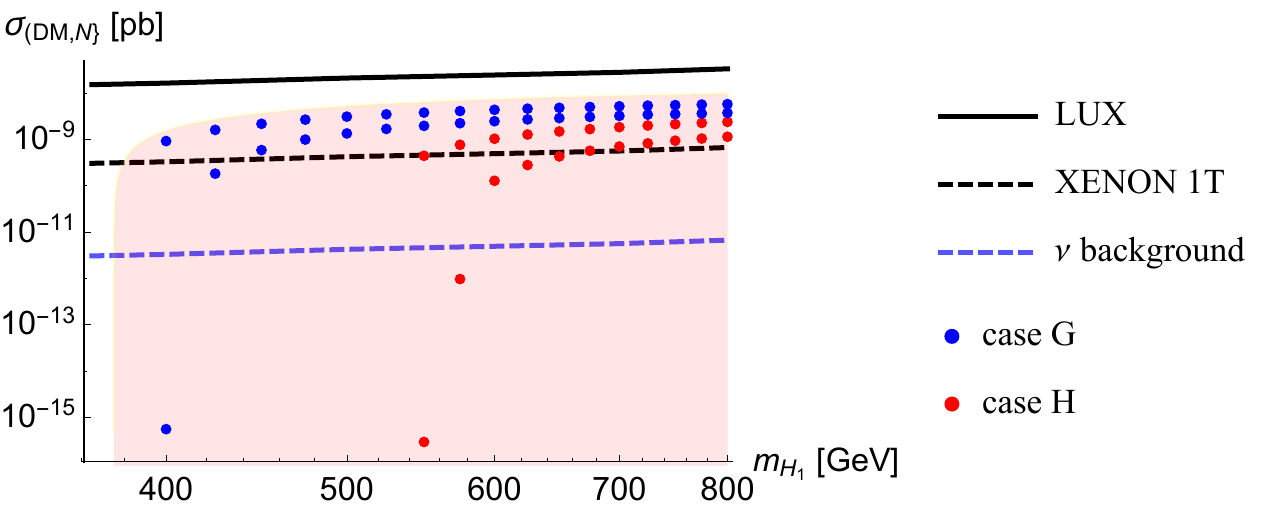}
\caption{The DM-nucleon scattering cross-section for the I(2+1)HDM in comparison with the direct detection limits set by LUX (black line) and projected XENON 1T (black dashed line). Coherent neutrino scattering limit is also shown (dashed blue line). The shaded region corresponds to points with relic density in agreement with Planck measurements; results for benchmark points for case G ($\Delta = 1 \gev, \delta = 1 \gev$, blue) and case H ($\Delta = 100 \gev, \delta = 1 \gev$) are presented.} 
\label{direct}
\end{figure}

These results are presented in Fig. \ref{direct} where the shaded region corresponds to the probed phase space of the I(2+1)HDM for various choices of $\Delta$ and $\delta$ (as shown in Fig. \ref{flag}), all of which have relic density in agreement with Planck measurements. Also, results for the benchmark points studied in section \ref{sec:annihilation} are presented explicitly. We found that the current experimental limits do not constrain the heavy DM mass region, neither for case H nor for case G, even though the Higgs-DM coupling is larger in the latter case.

Recall that for certain choices of $\Delta$ and $\delta$ one can obtain proper DM relic density for $g_{H_1 H_1 h} \approx 0$. This leads to a strongly suppressed scattering cross-section, which may not be detected as it lies within the coherent neutrino-nucleus scattering regime \cite{Aprile:2012zx}.

Fig. \ref{direct}  also shows a  limit from the future XENON1T experiment \cite{Anderson:2011bi}, with a proposed sensitivity of the order of $10^{-9}-10^{-10}$ pb (dashed black line). We expect the next generation of DM detectors, such as XENON1T, to be able to test a large portion of the parameter space of the heavy DM in the I(2+1)HDM for $m_{H_1} \lesssim 1$ TeV.

%\subsubsection*{Accidental cancellation in the I(2+1)HDM}
As a final note to this subsection, we should like to mention that a viable intermediate DM mass region, $m_W \lesssim m_{DM} \lesssim 160$ GeV, regarding relic density studies, has been found in the I(1+1)HDM. The correct relic density in this region is obtained due to cancellations between different diagrams contributing to DM annihilation into gauge bosons ($W^+W^-$ and $ZZ$). In \cite{LopezHonorez:2010tb} it was shown that this scenario is realised if the inert particles, in particular the charged scalar, are heavy enough, $\sim 300-500$ GeV. A relatively large DM-Higgs coupling is also required for the DM in this mass region to stay within relic density limits, however, this large Higgs-DM coupling is excluded by LUX. Similarly, we did not find any solutions in the medium mass region with viable relic density and in agreement with direct detection experiments in the I(2+1)HDM.

%%%%%%%%%%%%%%%%%%%%%%%%%%%%%%%%%%%%%%%%%%%%%%%%%%%%%%%%%%%%%%%%%%
\section{Heavy Higgs DM at the LHC in the I(2+1)HDM}\label{sec:LHC}

A scalar DM candidate is a stable particle with limited interactions with all SM particles and therefore it cannot be directly detected at the LHC. However, its presence can influence the detectable properties of SM particles. One way to ascertain the influence of DM candidate on the properties of a Higgs particle is to look at the Higgs invisible decays, $h\to SS$, where $S$ is a scalar DM candidate with mass below $m_{h}/2$. Invisible decays of the SM-like Higgs particle in the I(2+1)HDM were studied in \cite{Keus:2014jha, Keus:2014isa}, where limits for the mass of a light DM candidate combined with Planck limits for the relic density measurements provided constraints comparable or stronger than those from direct detection experiments.

Another strategy, useful for a heavy DM particle, is to look for a high $p_T$ monojet or a two jet/two lepton signal, accompanied by a large missing transverse energy $\Et$. The monojet signature in the I(2+1)HDM, $pp \to H_1 H_1 + \textrm{jet}$, corresponds to $h$ coupling to an invisible pair of DM particles (yielding the large $\Et$) with produced in association with an energetic quark or gluon jet. The following processes are considered in our analysis. 

\begin{enumerate}

\item $gg\to g H_1 H_1$ (Fig. \ref{Fig:gg}) via  a triple gluon and a $hgg$ effective vertex. Note, that the $hgg$ effective vertex in the I(2+1)HDM is the same as in the SM, as the Higgs production here is not modified by presence of additional scalar states. This is the dominant contribution to the monojet process, as the gluon fusion is an enhanced production mechanism for the Higgs particle.

\item $q\bar{q}\to g H_1 H_1$ (Fig. \ref{Fig:qbarq}), where $q = u,d,c,s,b$. The dominant contribution comes from the $s$-channel via the  $gq\bar{q}$ tree-level vertex and the $hgg$ effective coupling (Fig. \ref{Fig:qbarq}a).

\item $q g \to q H_1 H_1$ (Fig. \ref{Fig:qg}), where $q = u,d,c,s,b$. The dominant contributions here come from $gb\to H_1 H_1 b$ with the Higgs boson radiated off of the $b$ quark legs (Fig. \ref{Fig:qg}a) and $qg\to q H_1 H_1$ $t$-channel via a $gq\bar{q}$ tree-level vertex and the $hgg$ effective coupling (Fig. \ref{Fig:qg}b).

\end{enumerate}
Note, that all above processes contain the $h\to H_1H_1$ vertex, therefore a strong dependency on the $g_{H_1H_1h}$ coupling is expected, i.e., a significant difference between scenarios G and H, as discussed in the previous section.

For the studies of $pp \to H_1 H_1 + 2\textrm{jets}$ we have considered the Vector Boson Fusion (VBF) process of the form $q_i q_j \to H_1 H_1 q_k q_l$, with $q=u,d$ where a pair of DM particles (with large $\Et$) is produced by the neutral (Fig. \ref{diag:nvbf}) or charged (Fig. \ref{diag:chvbf}) VBF processes,  either directly or mediated by the Higgs particle or another neutral scalar.
 
We have also considered the Higgs-Strahlung (HS) processes of the form $q_i \bar{q_j} \to V^* H_1 H_1$ where a pair of DM particles is radiated off the $Z$ or $W^\pm$ boson leg (Fig. \ref{diag:nstr} and Fig. \ref{diag:chstr}, respectively), either directly or mediated by the Higgs particle or another neutral scalar.

Notice, that in the dijet searches only one diagram out of each set depends on the $g_{H_1 H_1 h}$, therefore we expect smaller differences between scenarios H and G than in the monojet searches. The strength of the other diagrams is set by the gauge interactions.

Also, it is important to stress that, given our initial choice of parameters, i.e. introducing the $k=1$ relation between the doublets (see Eq.(\ref{lambda-assumption})), we have limited the number of possible diagrams, because vertices of the type $Z H_i A_j$ and $W^\pm H_i H^\mp_j$ ($i \neq j$) are absent when $k=1$. Relaxing this initial assumption would in principle not only influence the evolution of DM relic density, but could also lead to a possibly stronger difference between scenarios G and H in the dijet analysis.

In the following subsections we present results for the monojet and dijet analysis, for the 14 TeV LHC. The following selections were used. \begin{enumerate}
\item 
For the monojet searches, we require the following cuts on the transverse momentum of the jet, $p_T^j$, and the pseudo-rapidity of the jet, $\eta^j$,
\be 
p_T^j > 20 \gev \quad \mbox{and} \quad |\eta^j| > 2.5
\ee

\item 
For the dijet searches, we require the following cuts on the invariant mass of the two jets, $M(j,j)$, and the difference between the pseudo-rapidity of the forward and backward jet,
\be 
M(j,j) > 700 \gev \quad \mbox{and} \quad  | \eta^j_f - \eta^j_b | > 4
\ee
\end{enumerate}

Calculations were done with the aid of LanHEP \cite{Semenov:2014rea} and CalcHEP \cite{Belyaev:2012qa} packages.

\subsection{Monojet results}

\begin{figure}[h!]
\centering
\includegraphics[scale=1]{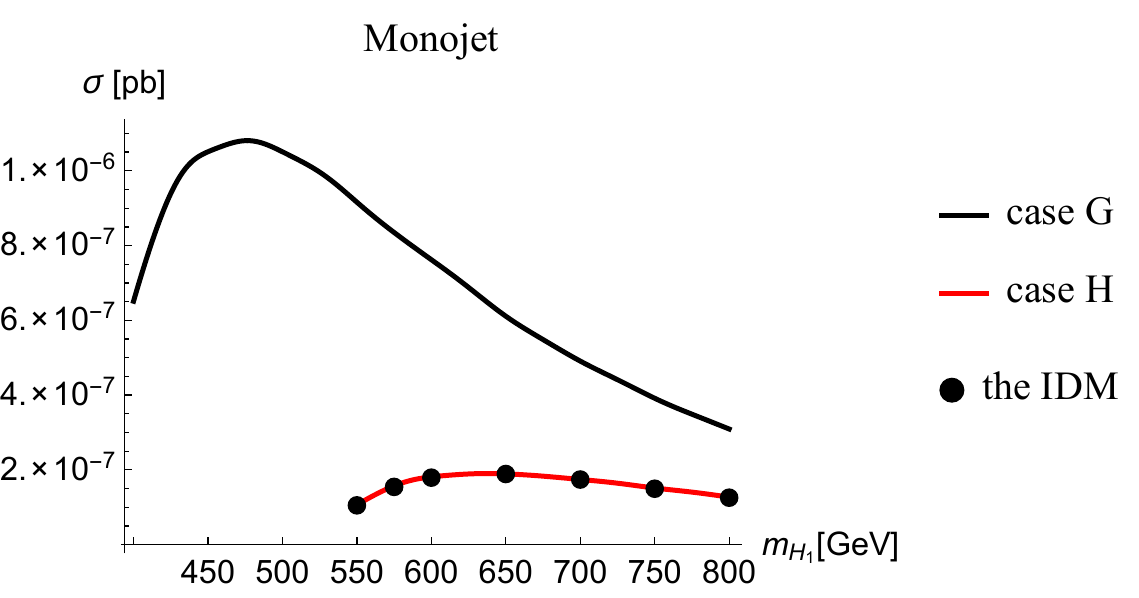}
\caption{Monojet searches for cases G and H. For comparison we also provide monojet searches for the I(1+1)HDM which resemble case H in the I(2+1)HDM as expected.}
\label{mono-fig}
\end{figure}

In Fig. \ref{mono-fig} results for monojet signals of scenarios G ($\delta_A=\delta_C=1 \gev, \Delta=1 \gev$) and H  ($\delta_A=\delta_C=1 \gev, \Delta=100\gev$) are shown. For comparison, we also present results for the I(1+1)HDM, with $\delta_A = \delta_C = 1$ from \cite{Krawczyk:2013jta}. The DM-Higgs coupling (defined as $\lambda_{345}$ in \cite{Krawczyk:2013jta}) is the same as $g_{H_1H_1 h}$ in scenario H for equal DM masses, therefore the monojet diagrams in case H and in I(1+1)HDM are identical. 
%Therefore, it is not surprising that the scenario H follows the the I(1+1)HDM case and cannot be distinguished from it. 

Scenario G, which corresponds to much larger Higgs-DM couplings compared to that of scenario H or the I(1+1)HDM, results in a significantly larger cross-section in the monojet process. Also the special features of the model are more visible in this process. For masses up to $m_{H_1} \approx 450 \gev$ we observe a rise in the cross-section connected to an opening of the phase space combined with an increasing Higgs-DM coupling. After that peak, the cross-section decreases with increasing DM mass regardless of the rising of $g_{h H_1 H_1}$.

Notice that for the lower masses, the difference between scenario G and H is significant, as every diagram involved in the monojet process contains  the $h H_1 H_1$ vertex, whose coupling differs significantly in cases G and H. Notice the low end of the allowed mass region in case G with a large cross-section for the mass region which is not even accessible by case H or the I(1+1)HDM. As the DM mass grows, results for both cases get closer together, stabilising for the very heavy mass region in the decoupling limit.

\subsection{Dijet VBF results}

\begin{figure}[h!]
\centering
\includegraphics[scale=1]{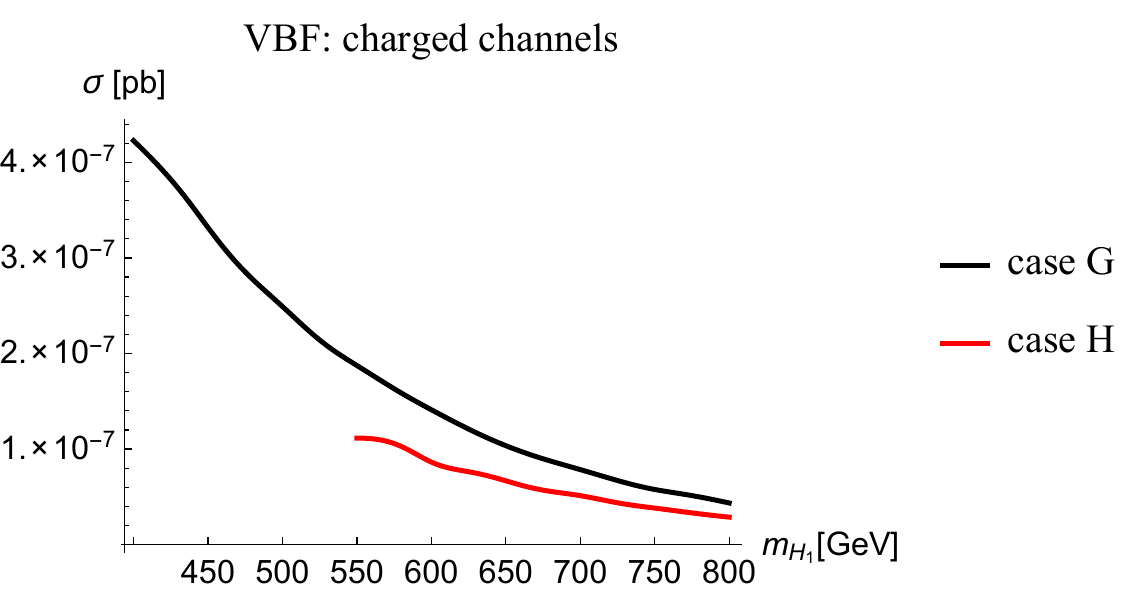}\\
\includegraphics[scale=1]{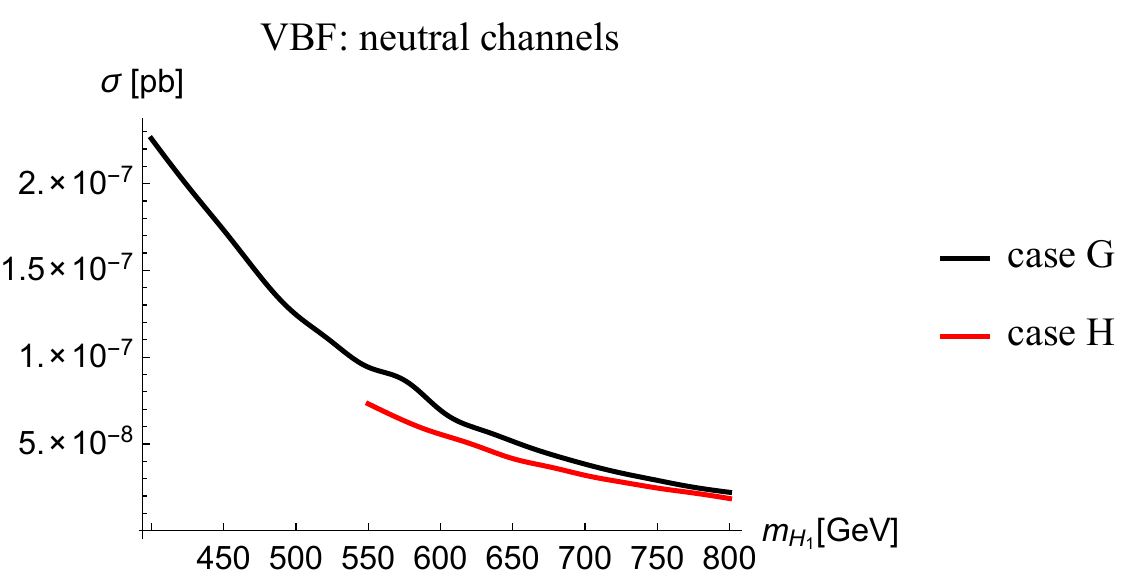}
\caption{The cross-section in dijet VBF processes in terms of the DM mass in charged (top) and neutral (bottom) processes.}
\label{dijet-VBF}
\end{figure}

Fig. \ref{dijet-VBF} presents values of the dijet cross-section for scenarios H and G in terms of the DM mass. The difference between cases H and G is less prominent compared to the monojet analysis, as only one of the involved diagrams in this process depends on the value of the Higgs-DM coupling (see Figs. \ref{diag:nvbf} and \ref{diag:chvbf} for diagrams involved in the neutral and charged VBF processes, respectively). We can still observe some differences in the lower mass range. The cross-section for  case G is generally larger (as is the $g_{h H_1 H_1}$ strength) than in case H. Also, charged channels have slightly larger cross-sections than the neutral ones, since the cross-section for producing the $W^\pm$ boson is larger than the cross-section for producing the $Z$ boson. For the heavier masses all results, for both scenarios G and H, as well as for charged and neutral channels, tend to converge.

\subsection{HS results}

HS signatures depend on the $W$ and $Z$ decay patterns. While at the LHC, leptonic signatures are preferred, hadronic ones are also possible. The latter potentially interfere with the VBF topologies, but the effect is small so that we can safely ignore it here.

\begin{figure}[h!]
\centering
\includegraphics[scale=1]{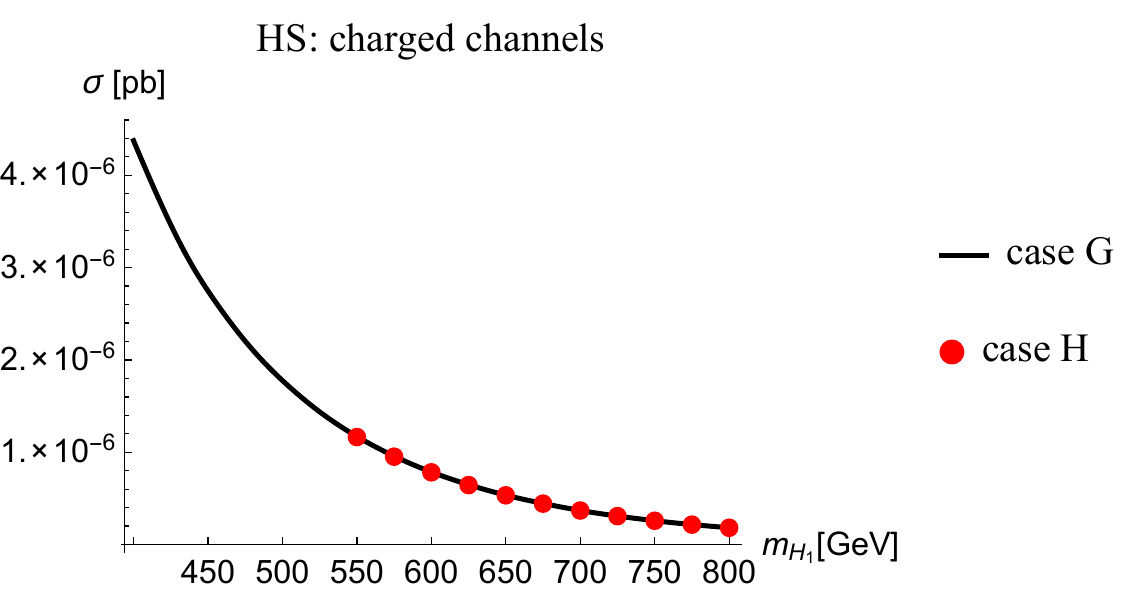}\\
\includegraphics[scale=1]{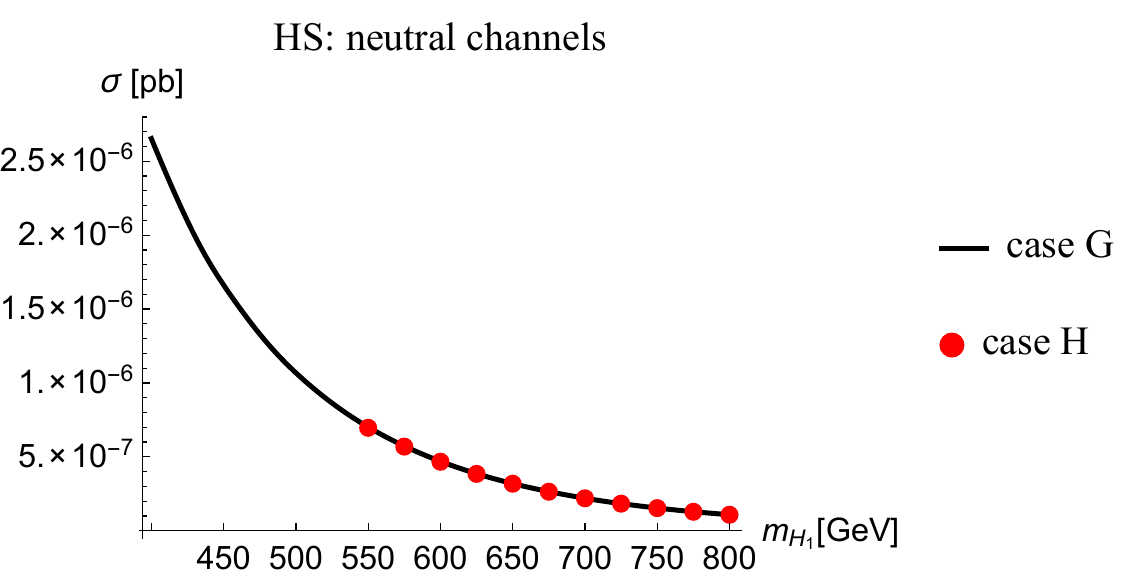}
\caption{The cross-section in HS processes in terms of the DM mass in charged (top) and neutral (bottom) processes.}
\label{dijet-HS}
\end{figure}

The results of the (on-shell) HS process cross-sections in terms of the DM mass are presented in Fig. \ref{dijet-HS}. It is clear that the general picture is different from the VBF studies. The largest cross-section again appears in the lower mass region where only case G can be realised. Similarly to the VBF case, the charged channels have larger cross-section compared to the neutral channels since the $W^\pm$ production in theses processes has a larger cross-section than the $Z$ boson production. All cross-sections in neutral and charged processes in both cases G and H converge to a similar value as $m_{DM}$ approaches very large values.

Note that the cross-section in both G and H are similar in the region of DM mass above $525 \gev$, i.e., where both cases can be realised (unlike in the VBF processes). This similarity is the result of the fact that the difference in the Higgs-DM coupling does not translate into a difference in the cross-section between cases G and H. To explain this similarity let us focus, e.g., on the neutral VBF and HS processes (Fig. \ref{diag:nvbf} and Fig. \ref{diag:nstr})\footnote{The same argument applies to comparing the charged VBF and HS processes.}. 

In the neutral VBF process, out of all the involved diagrams (Fig. \ref{diag:nvbf}a,b,c) there is only one diagram, Fig.  
 \ref{diag:nvbf}a, which depends on the Higgs-DM coupling. The cross-section of this diagram ($\sigma_h$) for a given $m_{DM}$ relative to the cross-section of all three diagrams involved ($\sigma_{tot}$) is $\sigma_{h} / \sigma_{tot} = 0.1445$ for case G and  $\sigma_{h}/\sigma_{tot} = 0.2416$ for case H. Recall that the main difference between cases G and H is that for a chosen $m_{DM}$ the Higgs-DM is larger in case G than in case H.
We conclude that the Higgs-mediated diagram plays a much more important role in case H than it does in case G. As a result, even though the $g_{hH_1H_1}$ coupling is much smaller in case H, the total cross-section does not fall far below the total cross-section in case G, which is depicted in Fig. \ref{dijet-VBF}.

Now, let us consider the HS neutral processes (Fig. \ref{diag:nstr}a,b,c). Again, only one diagram, Fig. \ref{diag:nstr}a, depends on the Higgs-DM coupling. Repeating the procedure above, we calculate the relative cross-section of this one diagram ($\sigma_h$) relative to the cross-section of all diagrams involved ($\sigma_{tot}$). For a given $m_{DM}$, we obtain $\sigma_{h}/\sigma_{tot} = 0.9736$ in case G and  $\sigma_{h}/\sigma_{tot} = 0.9708$ in case H. So, this diagram plays only a slight role in case H  compared to case G. 
We therefore conclude that the difference in $g_{hH_1H_1}$ coupling does not play an important role in distinguishing case G and H in the HS processes. Thus, we do not expect to see a difference between cases G and H, which is depicted in Fig. \ref{dijet-HS}, with the important exception that 
case G allows for the heavy Higgs DM mass to be below 525 GeV, making its cross-section significantly larger.

\section{Conclusion}
\label{conclusions}

We have calculated the relic density for heavy Higgs DM in the I(2+1)HDM and shown that the prospects for its
discovery at both DD experiments and the LHC are significantly enhanced as 
compared to the I(1+1)HDM, where the heavy Higgs DM particle must have a mass 
above about 525 GeV and is weakly coupled to the observed Higgs boson.
Adding a second inert Higgs doublet helps to make the heavy Higgs DM region accessible to 
both DD and the LHC, by either increasing its couplings to the observed Higgs or lowering its 
mass to $360 \gev \lesssim m_{DM}$, or both.
In particular we have presented LHC signatures of the I(2+1)HDM in the monojet, VBF (dijet) and HS processes and shown that the prospects for heavy Higgs DM discovery are significantly brighter for all channels.

In DD experiments, although 
the standard values of annihilation cross-section for the heavy Higgs DM masses in the I(2+1)HDM are well below current experimental exclusion limits for DM decaying into pairs of gauge bosons or fermions,
the prospects for a future DD discovery remain open due to the complementary nature of 
collider vs cosmological limits and the fact that the DD cross-sections are higher than in the 
I(1+1)HDM.

Turning to indirect detection signatures of the I(2+1)HDM, there is the possibility of internal bremsstrahlung in the processes of $H_1 H_1 \to W^+ W^- \gamma$, generated through the exchange of any of the two charged scalars $H^\pm_{1,2}$. It was shown that one can expect such signatures in the I(1+1)HDM  \cite{Garcia-Cely:2013zga}, mediated by a charged scalar in the $t$-channel, which would correspond to scenario H considered in the I(2+1)HDM. In principle, the signal could even be stronger for scenario G, as the scalar couplings are enhanced. 
%Note that in both scenarios for $k=1$ internal bremsstrahlung will be mediated only by one charged scalar, $H_1^\pm$, as there is no $H_1 H_2^\pm W^\mp$ vertex. Relaxing the initial assumption could lead to even stronger indirect detection signatures. 

Finally we comment on the observed $h \to \gamma \gamma$ channel where, 
in the I(1+1)HDM, only in the heavy DM mass region are both proper DM relic density and (minimal) enhancement in the $h \to \gamma \gamma$ channel realised. 
%For other mass regions, values of coupling that give good relic density will result in $\mu_{\gamma \gamma}$ below the SM value. In our work about the IDM we have shown that the significant enhancement (e.g. $\mu_{\gamma \gamma}\approx 1.3$) is possible only if the charged scalar, and therefore the DM candidate, is relatively light, $M_H \lesssim 150$ GeV. For heavier masses deviation from the SM value is smaller. 
By contrast, in the I(2+1)HDM there exist two charged scalars, $H_1^\pm$ and $H_2^\pm$, which contribute to the $h \to \gamma \gamma$ loop which may enhance the rate for a wide range of parameters.

%A guess based on the I(1+1)HDM studies -- to have a larger enhancement in $\mu_{\gamma \gamma}$ a larger value of $g_{H_1H_1g}$ is needed. This usually means, however, that the DM relic density is too small. In scenario G presented here good relic density corresponds to a larger couplings than in the I(1+1)HDM (or case H) and maybe it would be possible to have both good relic density and a slightly bigger $\mu_{\gamma \gamma}$. $H_{1,2}^\pm$ have the same coupling to Higgs in case $k=1$ (which is $\lambda_{23}v$), so there is no destructive interference between those two terms -- they will either both enhance the signal, or suppress  it. 

In conclusion, adding a second inert Higgs doublet significantly improves the prospects for observability of 
heavy Higgs dark matter in future experiments both underground and at the CERN LHC.

\section*{Acknowledgements}
SFK acknowledges partial support from the STFC Consolidated ST/J000396/ 1 and European Union FP7 ITN-INVISIBLES (Marie Curie Actions, PITN-GA-2011-289442). SM is financed in part through the NExT Institute and from the STFC Consolidated ST/ J000396/1. He also acknowledges the H2020-MSCA-RICE-2014 grant no. 645722 (NonMinimalHiggs).
VK's research is financially supported by the Academy of Finland project ``The Higgs Boson and the Cosmos'' and  project 267842.

\appendix
\section{Feynman rules in the simplified I(2+1)HDM}\label{Feyn-rules}
Here we present the Feynman rules of the model.
\bea
&& H^+_2 H^-_2 h, ~  H^+_1 H^-_1 h \qquad \qquad \lambda_{23} v \nonumber\\
&& H_1 H_1 h,~ H_2 H_2  h \qquad \qquad (\lambda_{23}+ \lambda'_{23} +2\lambda_{2}) \frac{v}{2} \nonumber\\
&& A_1 A_1 h,~ A_2 A_2  h \qquad \qquad (\lambda_{23} + \lambda'_{23} -2\lambda_{2} ) \frac{v}{2} \nonumber\\
&& H^+_2 H^-_2 \gamma,~ H^+_1 H^-_1  \gamma \qquad \qquad \frac{i}{2}(g\sin\theta_W + g'\cos\theta_W)(K+K')^\mu \nonumber\\
&& H^+_2 H^-_2 Z,~ H^+_1 H^-_1  Z \qquad \qquad \frac{i}{2}(g\cos\theta_W - g'\sin\theta_W)(K+K')^\mu \nonumber\\
&& H^\pm_1 H_1 W^\pm,~ H^\pm_2 H_2  W^\pm \qquad \qquad \frac{ig}{2}\cos(\theta_h -\theta_c)(K+K')^\mu \nonumber\\
&& H^\pm_2 H_1 W^\pm,~ H^\pm_1 H_2  W^\pm \qquad \qquad \frac{ig}{2}\sin(\theta_h -\theta_c)(K+K')^\mu \nonumber\\
&& H^\pm_1 A_1 W^\pm, ~H^\pm_2 A_2   W^\pm \qquad \qquad  \frac{g}{2} \cos(\theta_a -\theta_c)(K+K')^\mu \nonumber\\
&& H^\pm_2 A_1 W^\pm,~ H^\pm_1 A_2   W^\pm \qquad \qquad  \frac{g}{2} \sin(\theta_a -\theta_c)(K+K')^\mu \nonumber\\
&& H_1 A_1,~ H_2 A_2  Z \qquad \qquad   \frac{1}{2}(g\cos\theta_W + g'\sin\theta_W)\cos(\theta_h -\theta_a)(K+K')^\mu \nonumber\\
&& H_2 A_1, ~H_1 A_2  Z \qquad \qquad   \frac{1}{2}(g\cos\theta_W + g'\sin\theta_W)\sin(\theta_h -\theta_a)(K+K')^\mu \nonumber
\eea
where $K$ and $K'$ are the momenta of the associated particles in the decay channel and $\theta_W$ is the Weak mixing angle. The Yukawa couplings in the model are identical to those of the SM.

\section{Feynman diagrams for relic density calculations}
Here we present the DM (co)annihilation diagrams which play a role in our relic density studies.

\begin{minipage}{\linewidth}
      \centering
\begin{minipage}{1.0\linewidth}

          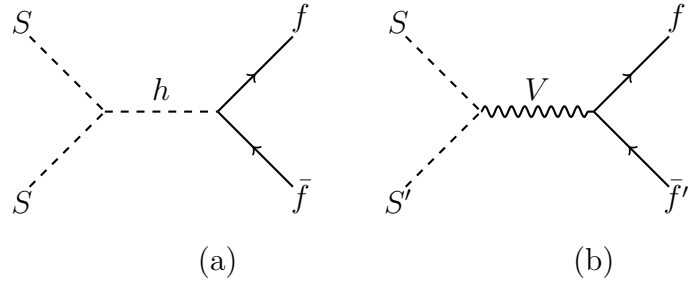
\begin{figure}[H]
          \centering
             \begin{tikzpicture}[thick,scale=1.0]
\draw[dashed] (0,0) -- node[black,above,xshift=-0.6cm,yshift=0.4cm] {$S$} (1,-1);
%\draw[antiparticle] (1,-1) -- node[black,above,yshift=-1.0cm,xshift=-0.6cm] {$\bar{q}$} (0,-2);
\draw[dashed] (0,-2) -- node[black,above,yshift=-1.0cm,xshift=-0.6cm] {$S$} (1,-1);
\draw[dashed] (1,-1) -- node[black,above,xshift=0.0cm,yshift=0.0cm] {$h$} (2.5,-1);
%\draw[dashed] (4,-1) -- node[black,above,yshift=-0.5cm,xshift=-0.2cm] {$\varphi$} (4,-2);
%\draw[particle] (4,-1) -- node[black,above,xshift=0.0cm,yshift=0.0cm] {$\chi_{b}$} (5,-1);
\draw[particle] (2.5,-1) -- node[black,above,xshift=0.6cm,yshift=0.4cm] {$f$} (3.5,0);
\draw[antiparticle] (2.5,-1) -- node[black,above,yshift=-1cm,xshift=0.6cm] {$\bar{f}$} (3.5,-2);

\node at (2.5,-3) {(a)};

\quad

\draw[dashed] (5,0) -- node[black,above,xshift=-0.6cm,yshift=0.4cm] {$S$} (6,-1);
%\draw[antiparticle] (1,-1) -- node[black,above,yshift=-1.0cm,xshift=-0.6cm] {$\bar{q}$} (0,-2);
\draw[dashed] (5,-2) -- node[black,above,yshift=-1.0cm,xshift=-0.6cm] {$S'$} (6,-1);
\draw[photon] (6,-1) -- node[black,above,xshift=0.0cm,yshift=0.0cm] {$V$} (7.5,-1);
%\draw[dashed] (4,-1) -- node[black,above,yshift=-0.5cm,xshift=-0.2cm] {$\varphi$} (4,-2);
%\draw[particle] (4,-1) -- node[black,above,xshift=0.0cm,yshift=0.0cm] {$\chi_{b}$} (5,-1);
\draw[particle] (7.5,-1) -- node[black,above,xshift=0.6cm,yshift=0.4cm] {$f$} (8.5,0);
\draw[antiparticle] (7.5,-1) -- node[black,above,yshift=-1cm,xshift=0.6cm] {$\bar{f'}$} (8.5,-2);

\node at (7.5,-3) {(b)};

%\draw\mypath;
\end{tikzpicture}
              \caption{For light DM masses the most important channel for the annihilation of DM  particles is the Higgs-mediated process (a). Coannhilation with other neutral scalars could have a significant effect on the relic density (b). }
\label{diag1}
          \end{figure}
      \end{minipage}
      \end{minipage}

\begin{minipage}{\linewidth}
      \centering
\begin{minipage}{1.0\linewidth}

          \begin{figure}[H]
          \centering
             \begin{tikzpicture}[thick,scale=1.0]
\draw[dashed] (0,0) -- node[black,above,xshift=-0.6cm,yshift=0.4cm] {$S$} (1,-1);
%\draw[antiparticle] (1,-1) -- node[black,above,yshift=-1.0cm,xshift=-0.6cm] {$\bar{q}$} (0,-2);
\draw[dashed] (0,-2) -- node[black,above,yshift=-1.0cm,xshift=-0.6cm] {$S'$} (1,-1);
\draw[dashed] (1,-1) -- node[black,above,xshift=0.0cm,yshift=0.0cm] {$h$} (2.5,-1);
%\draw[dashed] (4,-1) -- node[black,above,yshift=-0.5cm,xshift=-0.2cm] {$\varphi$} (4,-2);
%\draw[particle] (4,-1) -- node[black,above,xshift=0.0cm,yshift=0.0cm] {$\chi_{b}$} (5,-1);
\draw[photon] (2.5,-1) -- node[black,above,xshift=0.6cm,yshift=0.4cm] {$V,V^*$} (3.5,0);
\draw[photon] (2.5,-1) -- node[black,above,yshift=-1cm,xshift=0.6cm] {$V,V^*$} (3.5,-2);

\node at (2.5,-3) {(a)};

\quad

\draw[dashed] (7,0) -- node[black,above,xshift=-0.6cm,yshift=0.4cm] {$S$} (8,-1);
%\draw[antiparticle] (1,-1) -- node[black,above,yshift=-1.0cm,xshift=-0.6cm] {$\bar{q}$} (0,-2);
\draw[dashed] (7,-2) -- node[black,above,yshift=-1.0cm,xshift=-0.6cm] {$S$} (8,-1);
%\draw[photon] (6,-1) -- node[black,above,xshift=0.0cm,yshift=0.0cm] {$V$} (7.5,-1);
%\draw[dashed] (4,-1) -- node[black,above,yshift=-0.5cm,xshift=-0.2cm] {$\varphi$} (4,-2);
%\draw[particle] (4,-1) -- node[black,above,xshift=0.0cm,yshift=0.0cm] {$\chi_{b}$} (5,-1);
\draw[photon] (8,-1) -- node[black,above,xshift=0.6cm,yshift=0.4cm] {$V,V^*$} (9,0);
\draw[photon] (8,-1) -- node[black,above,yshift=-1cm,xshift=0.6cm] {$V,V^*$} (9,-2);

\node at (7.5,-3) {(b)};

%\draw\mypath;
\end{tikzpicture}
              \caption{For heavy DM masses, the processes involving one or two virtual gauge bosons, also contribute to the total annihilation cross-section and affect the relic density.}
\label{diag2}
          \end{figure}
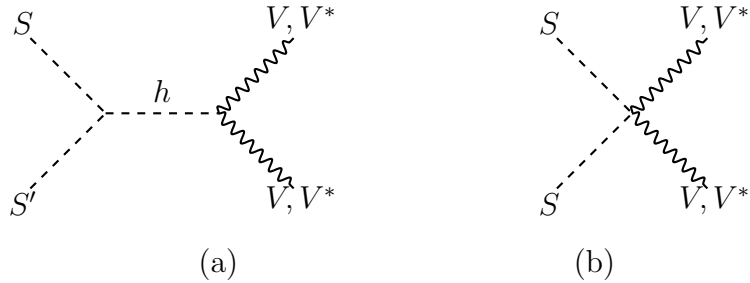
      \end{minipage}
      \end{minipage}

\begin{minipage}{\linewidth}
      \centering
\begin{minipage}{1.0\linewidth}

          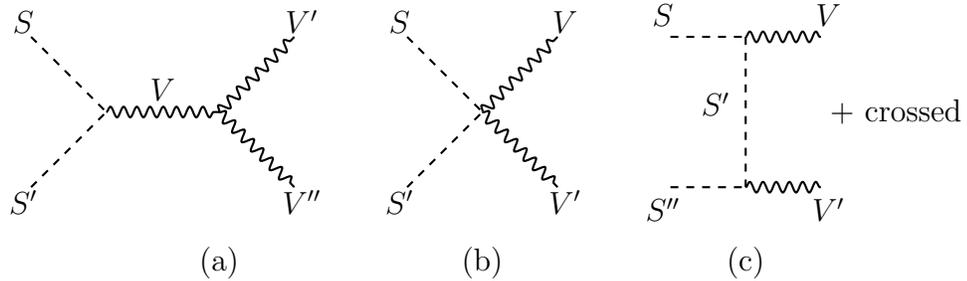
\begin{figure}[H]
          \centering
             \begin{tikzpicture}[thick,scale=1.0]
\draw[dashed] (0,0) -- node[black,above,xshift=-0.6cm,yshift=0.4cm] {$S$} (1,-1);
%\draw[antiparticle] (1,-1) -- node[black,above,yshift=-1.0cm,xshift=-0.6cm] {$\bar{q}$} (0,-2);
\draw[dashed] (0,-2) -- node[black,above,yshift=-1.0cm,xshift=-0.6cm] {$S'$} (1,-1);
\draw[photon] (1,-1) -- node[black,above,xshift=0.0cm,yshift=0.0cm] {$V$} (2.5,-1);
%\draw[dashed] (4,-1) -- node[black,above,yshift=-0.5cm,xshift=-0.2cm] {$\varphi$} (4,-2);
%\draw[particle] (4,-1) -- node[black,above,xshift=0.0cm,yshift=0.0cm] {$\chi_{b}$} (5,-1);
\draw[photon] (2.5,-1) -- node[black,above,xshift=0.6cm,yshift=0.4cm] {$V'$} (3.5,0);
\draw[photon] (2.5,-1) -- node[black,above,yshift=-1cm,xshift=0.6cm] {$V''$} (3.5,-2);

\node at (2.5,-3) {(a)};

\quad

\draw[dashed] (5,0) -- node[black,above,xshift=-0.6cm,yshift=0.4cm] {$S$} (6,-1);
%\draw[antiparticle] (1,-1) -- node[black,above,yshift=-1.0cm,xshift=-0.6cm] {$\bar{q}$} (0,-2);
\draw[dashed] (5,-2) -- node[black,above,yshift=-1.0cm,xshift=-0.6cm] {$S'$} (6,-1);
%\draw[photon] (6,-1) -- node[black,above,xshift=0.0cm,yshift=0.0cm] {$V$} (7.5,-1);
%\draw[dashed] (4,-1) -- node[black,above,yshift=-0.5cm,xshift=-0.2cm] {$\varphi$} (4,-2);
%\draw[particle] (4,-1) -- node[black,above,xshift=0.0cm,yshift=0.0cm] {$\chi_{b}$} (5,-1);
\draw[photon] (6,-1) -- node[black,above,xshift=0.6cm,yshift=0.4cm] {$V$} (7,0);
\draw[photon] (6,-1) -- node[black,above,yshift=-1cm,xshift=0.6cm] {$V'$} (7,-2);

\node at (6,-3) {(b)};

\quad

\draw[dashed] (8.5,0) -- node[black,above,xshift=-0.6cm,yshift=0.0cm] {$S$} (9.5,0);
%\draw[antiparticle] (1,-1) -- node[black,above,yshift=-1.0cm,xshift=-0.6cm] {$\bar{q}$} (0,-2);
\draw[dashed] (8.5,-2) -- node[black,above,yshift=-0.6cm,xshift=-0.6cm] {$S''$} (9.5,-2);
\draw[dashed] (9.5,0) -- node[black,above,xshift=-0.4cm,yshift=-0.2cm] {$S'$} (9.5,-2);
%\draw[dashed] (4,-1) -- node[black,above,yshift=-0.5cm,xshift=-0.2cm] {$\varphi$} (4,-2);
%\draw[particle] (4,-1) -- node[black,above,xshift=0.0cm,yshift=0.0cm] {$\chi_{b}$} (5,-1);
\draw[photon] (9.5,0) -- node[black,above,xshift=0.6cm,yshift=0.0cm] {$V$} (10.5,0);
\draw[photon] (9.5,-2) -- node[black,above,yshift=-0.6cm,xshift=0.6cm] {$V'$} (10.5,-2);

\node at (11.5,-1) {+ crossed};

\node at (9.5,-3) {(c)};

%\draw\mypath;
\end{tikzpicture}
              \caption{Heavy DM (co)annihilation diagrams with pure gauge boson final states.}
\label{diag3}
          \end{figure}
      \end{minipage}
      \end{minipage}

\begin{minipage}{\linewidth}
      \centering
\begin{minipage}{1.0\linewidth}

          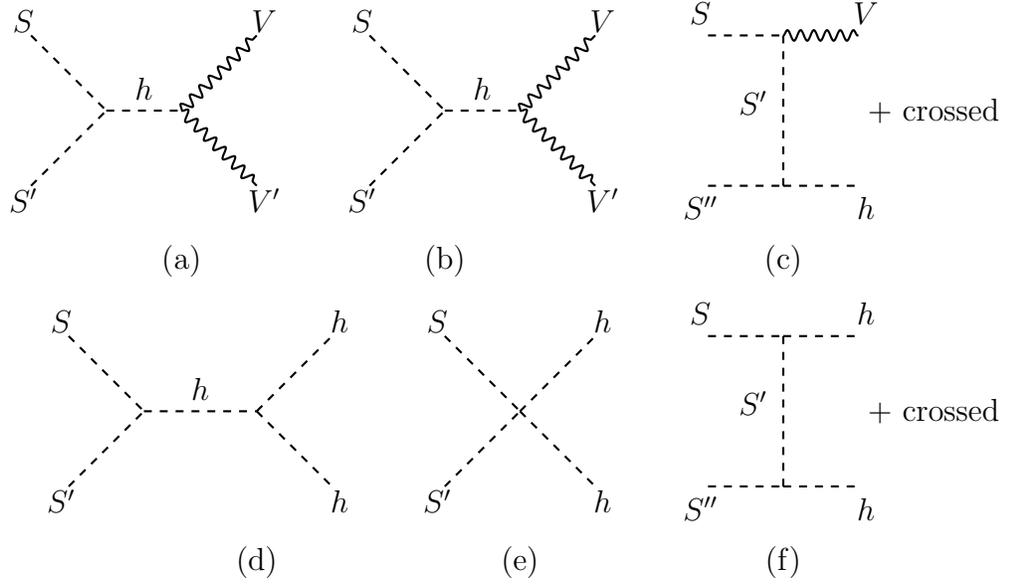
\begin{figure}[H]
          \centering
          
                       \begin{tikzpicture}[thick,scale=1.0]

\draw[dashed] (0,0) -- node[black,above,xshift=-0.6cm,yshift=0.4cm] {$S$} (1,-1);
%\draw[antiparticle] (1,-1) -- node[black,above,yshift=-1.0cm,xshift=-0.6cm] {$\bar{q}$} (0,-2);
\draw[dashed] (0,-2) -- node[black,above,yshift=-1.0cm,xshift=-0.6cm] {$S'$} (1,-1);
\draw[dashed] (1,-1) -- node[black,above,xshift=0.0cm,yshift=0.0cm] {$h$} (2,-1);
%\draw[dashed] (4,-1) -- node[black,above,yshift=-0.5cm,xshift=-0.2cm] {$\varphi$} (4,-2);
%\draw[particle] (4,-1) -- node[black,above,xshift=0.0cm,yshift=0.0cm] {$\chi_{b}$} (5,-1);
\draw[photon] (2,-1) -- node[black,above,xshift=0.6cm,yshift=0.4cm] {$V$} (3,0);
\draw[photon] (2,-1) -- node[black,above,yshift=-1cm,xshift=0.6cm] {$V'$} (3,-2);

\node at (2,-3) {(a)};

\quad

\draw[dashed] (4.5,0) -- node[black,above,xshift=-0.6cm,yshift=0.4cm] {$S$} (5.5,-1);
%\draw[antiparticle] (1,-1) -- node[black,above,yshift=-1.0cm,xshift=-0.6cm] {$\bar{q}$} (0,-2);
\draw[dashed] (4.5,-2) -- node[black,above,yshift=-1.0cm,xshift=-0.6cm] {$S'$} (5.5,-1);
\draw[dashed] (5.5,-1) -- node[black,above,xshift=0.0cm,yshift=0.0cm] {$h$} (6.5,-1);
%\draw[photon] (6,-1) -- node[black,above,xshift=0.0cm,yshift=0.0cm] {$V$} (7.5,-1);
%\draw[dashed] (4,-1) -- node[black,above,yshift=-0.5cm,xshift=-0.2cm] {$\varphi$} (4,-2);
%\draw[particle] (4,-1) -- node[black,above,xshift=0.0cm,yshift=0.0cm] {$\chi_{b}$} (5,-1);
\draw[photon] (6.5,-1) -- node[black,above,xshift=0.6cm,yshift=0.4cm] {$V$} (7.5,0);
\draw[photon] (6.5,-1) -- node[black,above,yshift=-1cm,xshift=0.6cm] {$V'$} (7.5,-2);

\node at (5.5,-3) {(b)};

\quad
\hspace{0.5cm}

\draw[dashed] (8.5,0) -- node[black,above,xshift=-0.6cm,yshift=0.0cm] {$S$} (9.5,0);
%\draw[antiparticle] (1,-1) -- node[black,above,yshift=-1.0cm,xshift=-0.6cm] {$\bar{q}$} (0,-2);
\draw[dashed] (8.5,-2) -- node[black,above,yshift=-0.6cm,xshift=-0.6cm] {$S''$} (9.5,-2);
\draw[dashed] (9.5,0) -- node[black,above,xshift=-0.4cm,yshift=-0.2cm] {$S'$} (9.5,-2);
%\draw[dashed] (4,-1) -- node[black,above,yshift=-0.5cm,xshift=-0.2cm] {$\varphi$} (4,-2);
%\draw[particle] (4,-1) -- node[black,above,xshift=0.0cm,yshift=0.0cm] {$\chi_{b}$} (5,-1);
\draw[photon] (9.5,0) -- node[black,above,xshift=0.6cm,yshift=0.0cm] {$V$} (10.5,0);
\draw[dashed] (9.5,-2) -- node[black,above,yshift=-0.6cm,xshift=0.6cm] {$h$} (10.5,-2);

\node at (11.5,-1) {+ crossed};

\node at (9.5,-3) {(c)};

\draw[dashed] (0,-4) -- node[black,above,xshift=-0.6cm,yshift=0.4cm] {$S$} (1,-5);
%\draw[antiparticle] (1,-1) -- node[black,above,yshift=-1.0cm,xshift=-0.6cm] {$\bar{q}$} (0,-2);
\draw[dashed] (0,-6) -- node[black,above,yshift=-1.0cm,xshift=-0.6cm] {$S'$} (1,-5);
\draw[dashed] (1,-5) -- node[black,above,xshift=0.0cm,yshift=0.0cm] {$h$} (2.5,-5);
%\draw[dashed] (4,-1) -- node[black,above,yshift=-0.5cm,xshift=-0.2cm] {$\varphi$} (4,-2);
%\draw[particle] (4,-1) -- node[black,above,xshift=0.0cm,yshift=0.0cm] {$\chi_{b}$} (5,-1);
\draw[dashed] (2.5,-5) -- node[black,above,xshift=0.6cm,yshift=0.4cm] {$h$} (3.5,-4);
\draw[dashed] (2.5,-5) -- node[black,above,yshift=-1cm,xshift=0.6cm] {$h$} (3.5,-6);

\node at (2.5,-7) {(d)};

\quad

\draw[dashed] (5,-4) -- node[black,above,xshift=-0.6cm,yshift=0.4cm] {$S$} (6,-5);
%\draw[antiparticle] (1,-1) -- node[black,above,yshift=-1.0cm,xshift=-0.6cm] {$\bar{q}$} (0,-2);
\draw[dashed] (5,-6) -- node[black,above,yshift=-1.0cm,xshift=-0.6cm] {$S'$} (6,-5);
%\draw[photon] (6,-1) -- node[black,above,xshift=0.0cm,yshift=0.0cm] {$V$} (7.5,-5);
%\draw[dashed] (4,-1) -- node[black,above,yshift=-0.5cm,xshift=-0.2cm] {$\varphi$} (4,-2);
%\draw[particle] (4,-1) -- node[black,above,xshift=0.0cm,yshift=0.0cm] {$\chi_{b}$} (5,-1);
\draw[dashed] (6,-5) -- node[black,above,xshift=0.6cm,yshift=0.4cm] {$h$} (7,-4);
\draw[dashed] (6,-5) -- node[black,above,yshift=-1cm,xshift=0.6cm] {$h$} (7,-6);

\node at (6,-7) {(e)};

\quad

\draw[dashed] (8.5,-4) -- node[black,above,xshift=-0.6cm,yshift=0.0cm] {$S$} (9.5,-4);
%\draw[antiparticle] (1,-1) -- node[black,above,yshift=-1.0cm,xshift=-0.6cm] {$\bar{q}$} (0,-2);
\draw[dashed] (8.5,-6) -- node[black,above,yshift=-0.6cm,xshift=-0.6cm] {$S''$} (9.5,-6);
\draw[dashed] (9.5,-4) -- node[black,above,xshift=-0.4cm,yshift=-0.2cm] {$S'$} (9.5,-6);
%\draw[dashed] (4,-1) -- node[black,above,yshift=-0.5cm,xshift=-0.2cm] {$\varphi$} (4,-2);
%\draw[particle] (4,-1) -- node[black,above,xshift=0.0cm,yshift=0.0cm] {$\chi_{b}$} (5,-1);
\draw[dashed] (9.5,-4) -- node[black,above,xshift=0.6cm,yshift=0.0cm] {$h$} (10.5,-4);
\draw[dashed] (9.5,-6) -- node[black,above,yshift=-0.6cm,xshift=0.6cm] {$h$} (10.5,-6);

\node at (11.5,-5) {+ crossed};

\node at (9.5,-7) {(f)};

%\draw\mypath;
\end{tikzpicture}
              \caption{Heavy DM (co)annhilation channels involving the SM-like Higgs boson.}
\label{diag4}
          \end{figure}
      \end{minipage}
      \end{minipage}

\vspace{5mm}

\section{Feynman diagrams for the LHC analysis}
Here we present the DM (co)annihilation diagrams which play a role in our LHC studies.

\subsection{Diagrams with monojet final states}

\begin{minipage}{\linewidth}
      \centering
\begin{minipage}{1.0\linewidth}

          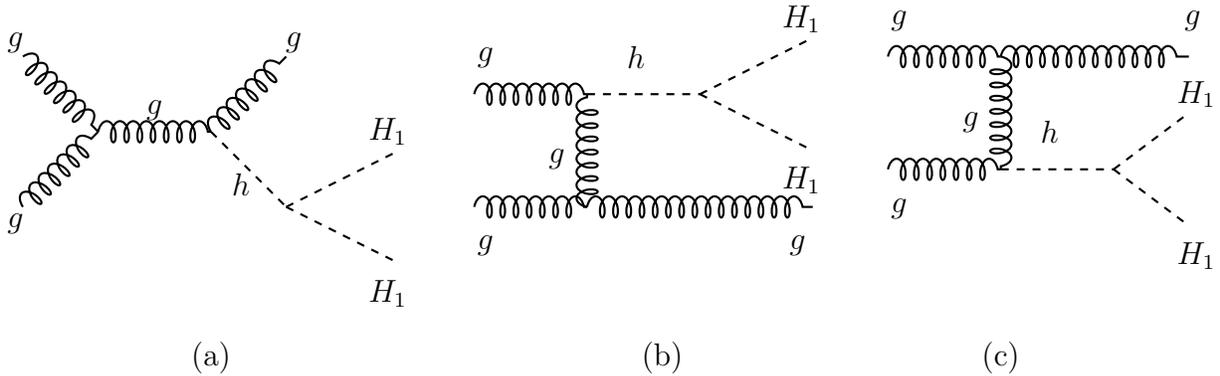
\begin{figure}[H]
          
             \begin{tikzpicture}[thick,scale=1.0]
\draw[gluon] (0,0) -- node[black,above,xshift=-0.6cm,yshift=0.4cm] {$g$} (1,-1);
%\draw[antiparticle] (1,-1) -- node[black,above,yshift=-1.0cm,xshift=-0.6cm] {$\bar{q}$} (0,-2);
\draw[gluon] (0,-2) -- node[black,above,yshift=-1.0cm,xshift=-0.6cm] {$g$} (1,-1);
\draw[gluon] (1,-1) -- node[black,above,xshift=0.0cm,yshift=0.0cm] {$g$} (2.5,-1);
%\draw[dashed] (4,-1) -- node[black,above,yshift=-0.5cm,xshift=-0.2cm] {$\varphi$} (4,-2);
%\draw[particle] (4,-1) -- node[black,above,xshift=0.0cm,yshift=0.0cm] {$\chi_{b}$} (5,-1);
\draw[gluon] (2.5,-1) -- node[black,above,xshift=0.6cm,yshift=0.4cm] {$g$} (3.5,0);
\draw[dashed] (2.5,-1) -- node[black,above,yshift=-0.5cm,xshift=-0.1cm] {$h$} (3.5,-2);
\draw[dashed] (3.5,-2) -- node[black,above,yshift=0.3cm,xshift=0.6cm] {$H_1$} (5,-1.25);
\draw[dashed] (3.5,-2) -- node[black,above,yshift=-1.1cm,xshift=0.6cm] {$H_1$} (5,-2.75);

\node at (2.5,-4) {(a)};

\quad

\draw[gluon] (6,-0.5) -- node[black,above,xshift=-0.6cm,yshift=0.2cm] {$g$} (7.5,-0.5);
\draw[gluon] (6,-2) -- node[black,above,yshift=-0.8cm,xshift=-0.6cm] {$g$} (7.5,-2);
\draw[gluon] (7.5,-2) -- node[black,above,xshift=-0.4cm,yshift=-0.4cm] {$g$} (7.5,-0.5);
\draw[gluon] (7.5,-2) -- node[black,above,xshift=1.3cm,yshift=-0.8cm] {$g$} (10.5,-2);
\draw[dashed] (7.5,-0.5) -- node[black,above,yshift=0.2cm,xshift=-0.1cm] {$h$} (9,-0.5);
\draw[dashed] (9,-0.5) -- node[black,above,yshift=0.3cm,xshift=0.6cm] {$H_1$} (10.5,0.25);
\draw[dashed] (9,-0.5) -- node[black,above,yshift=-1.1cm,xshift=0.6cm] {$H_1$} (10.5,-1.25);

\node at (8.5,-4) {(b)};

\draw[gluon] (11.5,0) -- node[black,above,xshift=-0.6cm,yshift=0.2cm] {$g$} (13,0);
\draw[gluon] (11.5,-1.5) -- node[black,above,yshift=-0.8cm,xshift=-0.6cm] {$g$} (13,-1.5);
\draw[gluon] (13,0) -- node[black,above,xshift=-0.4cm,yshift=-0.4cm] {$g$} (13,-1.5);
\draw[gluon] (13,0) -- node[black,above,xshift=1.3cm,yshift=0.2cm] {$g$} (15.5,0);
\draw[dashed] (13,-1.5) -- node[black,above,yshift=0.2cm,xshift=-0.1cm] {$h$} (14.5,-1.5);
\draw[dashed] (14.5,-1.5) -- node[black,above,yshift=0.3cm,xshift=0.6cm] {$H_1$} (15.5,-0.75);
\draw[dashed] (14.5,-1.5) -- node[black,above,yshift=-1.1cm,xshift=0.6cm] {$H_1$} (15.5,-2.25);

\node at (13,-4) {(c)};

%\draw\mypath;
\end{tikzpicture}
\caption{Relevant monojet diagrams with initial gluon states ($gg\to hg \to g H_1 H_1$) \\
containing triple gluon vertex and an effective $ggh$ vertex.}
\label{Fig:gg}
          \end{figure}
      \end{minipage}
      \end{minipage}

\begin{minipage}{\linewidth}
\centering
\begin{minipage}{1.0\linewidth}

          \begin{figure}[H]
          
             \begin{tikzpicture}[thick,scale=1.0]
\draw[particle] (0,0) -- node[black,above,xshift=-0.6cm,yshift=0.4cm] {$q$} (1,-1);
%\draw[antiparticle] (1,-1) -- node[black,above,yshift=-1.0cm,xshift=-0.6cm] {$\bar{q}$} (0,-2);
\draw[antiparticle] (0,-2) -- node[black,above,yshift=-1.0cm,xshift=-0.6cm] {$\bar{q}$} (1,-1);
\draw[gluon] (1,-1) -- node[black,above,xshift=0.0cm,yshift=0.0cm] {$g$} (2.5,-1);
%\draw[dashed] (4,-1) -- node[black,above,yshift=-0.5cm,xshift=-0.2cm] {$\varphi$} (4,-2);
%\draw[particle] (4,-1) -- node[black,above,xshift=0.0cm,yshift=0.0cm] {$\chi_{b}$} (5,-1);
\draw[gluon] (2.5,-1) -- node[black,above,xshift=0.6cm,yshift=0.4cm] {$g$} (3.5,0);
\draw[dashed] (2.5,-1) -- node[black,above,yshift=-0.5cm,xshift=-0.1cm] {$h$} (3.5,-2);
\draw[dashed] (3.5,-2) -- node[black,above,yshift=0.3cm,xshift=0.6cm] {$H_1$} (5,-1.25);
\draw[dashed] (3.5,-2) -- node[black,above,yshift=-1.1cm,xshift=0.6cm] {$H_1$} (5,-2.75);

\node at (2.5,-4) {(a)};

\quad

\draw[particle] (6,-0.5) -- node[black,above,xshift=-0.6cm,yshift=0.2cm] {$q$} (7.5,-0.5);
\draw[antiparticle] (6,-2) -- node[black,above,yshift=-0.8cm,xshift=-0.6cm] {$\bar{q}$} (7.5,-2);
\draw[particle] (7.5,-0.5) -- node[black,above,xshift=-0.4cm,yshift=-0.4cm] {$q$} (7.5,-2);
\draw[gluon] (7.5,-2) -- node[black,above,xshift=1.3cm,yshift=-0.8cm] {$g$} (10.5,-2);
\draw[dashed] (7.5,-0.5) -- node[black,above,yshift=0.2cm,xshift=-0.1cm] {$h$} (9,-0.5);
\draw[dashed] (9,-0.5) -- node[black,above,yshift=0.3cm,xshift=0.6cm] {$H_1$} (10.5,0.25);
\draw[dashed] (9,-0.5) -- node[black,above,yshift=-1.1cm,xshift=0.6cm] {$H_1$} (10.5,-1.25);

\node at (8.5,-4) {(b)};

\draw[particle] (11.5,0) -- node[black,above,xshift=-0.6cm,yshift=0.2cm] {$q$} (13,0);
\draw[antiparticle] (11.5,-1.5) -- node[black,above,yshift=-0.8cm,xshift=-0.6cm] {$\bar{q}$} (13,-1.5);
\draw[particle] (13,0) -- node[black,above,xshift=-0.4cm,yshift=-0.4cm] {$q$} (13,-1.5);
\draw[gluon] (13,0) -- node[black,above,xshift=1.3cm,yshift=0.2cm] {$g$} (15.5,0);
\draw[dashed] (13,-1.5) -- node[black,above,yshift=0.2cm,xshift=-0.1cm] {$h$} (14.5,-1.5);
\draw[dashed] (14.5,-1.5) -- node[black,above,yshift=0.3cm,xshift=0.6cm] {$H_1$} (15.5,-0.75);
\draw[dashed] (14.5,-1.5) -- node[black,above,yshift=-1.1cm,xshift=0.6cm] {$H_1$} (15.5,-2.25);

\node at (13,-4) {(c)};

%\draw\mypath;
\end{tikzpicture}
              \caption{Relevant monojet diagrams with initial quark states ($q\bar{q}\to g H_1 H_1$ + diagrams with initial particles reversed) containing $ggh$ effective vertex, where $q = u,d,c,s,b$.}
\label{Fig:qbarq}
          \end{figure}
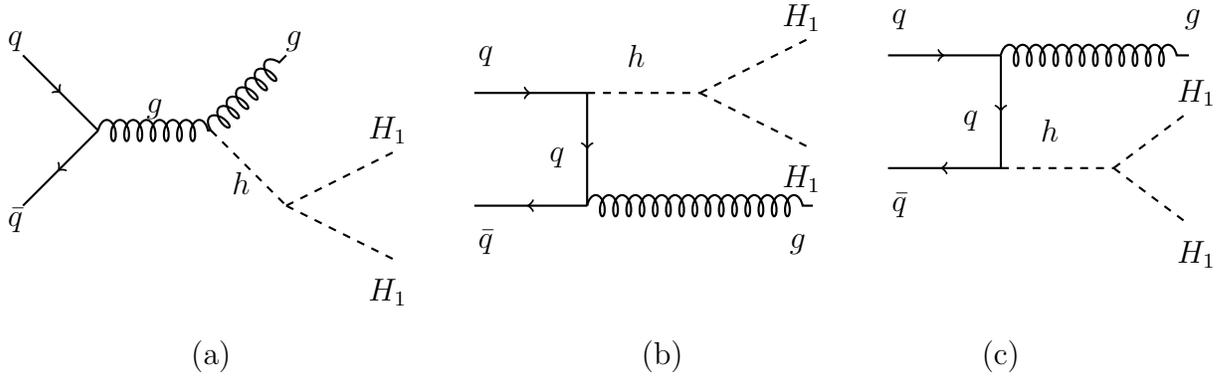
      \end{minipage}
      \end{minipage}

\begin{minipage}{\linewidth}
    \centering
\begin{minipage}{1.0\linewidth}

          \begin{figure}[H]
          
             \begin{tikzpicture}[thick,scale=1.0]
\draw[particle] (0,0) -- node[black,above,xshift=-0.6cm,yshift=0.4cm] {$q$} (1,-1);
\draw[gluon] (1,-1) -- node[black,above,yshift=-1.0cm,xshift=-0.6cm] {$g$} (0,-2);
\draw[particle] (1,-1) -- node[black,above,xshift=0.0cm,yshift=0.0cm] {$q$} (2.5,-1);
%\draw[dashed] (4,-1) -- node[black,above,yshift=-0.5cm,xshift=-0.2cm] {$\varphi$} (4,-2);
%\draw[particle] (4,-1) -- node[black,above,xshift=0.0cm,yshift=0.0cm] {$\chi_{b}$} (5,-1);
\draw[particle] (2.5,-1) -- node[black,above,xshift=0.6cm,yshift=0.4cm] {$q$} (3.5,0);
\draw[dashed] (2.5,-1) -- node[black,above,yshift=-0.5cm,xshift=-0.1cm] {$h$} (3.5,-2);
\draw[dashed] (3.5,-2) -- node[black,above,yshift=0.3cm,xshift=0.6cm] {$H_1$} (5,-1.25);
\draw[dashed] (3.5,-2) -- node[black,above,yshift=-1.1cm,xshift=0.6cm] {$H_1$} (5,-2.75);

\node at (2.5,-4) {(a)};

\quad

\draw[particle] (6,-0.5) -- node[black,above,xshift=-0.6cm,yshift=0.2cm] {$q$} (7.5,-0.5);
\draw[gluon] (6,-2) -- node[black,above,yshift=-0.8cm,xshift=-0.6cm] {$g$} (7.5,-2);
\draw[particle] (7.5,-0.5) -- node[black,above,xshift=-0.4cm,yshift=-0.4cm] {$q$} (7.5,-2);
\draw[particle] (7.5,-2) -- node[black,above,xshift=1.3cm,yshift=-0.8cm] {$q$} (10.5,-2);
\draw[dashed] (7.5,-0.5) -- node[black,above,yshift=0.2cm,xshift=-0.1cm] {$h$} (9,-0.5);
\draw[dashed] (9,-0.5) -- node[black,above,yshift=0.3cm,xshift=0.6cm] {$H_1$} (10.5,0.25);
\draw[dashed] (9,-0.5) -- node[black,above,yshift=-1.1cm,xshift=0.6cm] {$H_1$} (10.5,-1.25);

\node at (8.5,-4) {(b)};

\draw[particle] (11.5,0) -- node[black,above,xshift=-0.6cm,yshift=0.2cm] {$q$} (13,0);
\draw[gluon] (11.5,-1.5) -- node[black,above,yshift=-0.8cm,xshift=-0.6cm] {$g$} (13,-1.5);
\draw[gluon] (13,0) -- node[black,above,xshift=-0.4cm,yshift=-0.4cm] {$g$} (13,-1.5);
\draw[particle] (13,0) -- node[black,above,xshift=1.3cm,yshift=0.2cm] {$q$} (15.5,0);
\draw[dashed] (13,-1.5) -- node[black,above,yshift=0.2cm,xshift=-0.1cm] {$h$} (14.5,-1.5);
\draw[dashed] (14.5,-1.5) -- node[black,above,yshift=0.3cm,xshift=0.6cm] {$H_1$} (15.5,-0.75);
\draw[dashed] (14.5,-1.5) -- node[black,above,yshift=-1.1cm,xshift=0.6cm] {$H_1$} (15.5,-2.25);

\node at (13,-4) {(c)};

%\draw\mypath;
\end{tikzpicture}
\caption{Relevant monojet diagrams with initial quark and gluon states ($qg\to q H_1 H_1$ + equivalent $\bar{q}g\to \bar{q} H_1 H_1$ diagrams + diagrams with initial particles reversed) containing  $ggh$ effective vertex, where $q = u,d,c,s,b$.}
\label{Fig:qg}
          \end{figure}
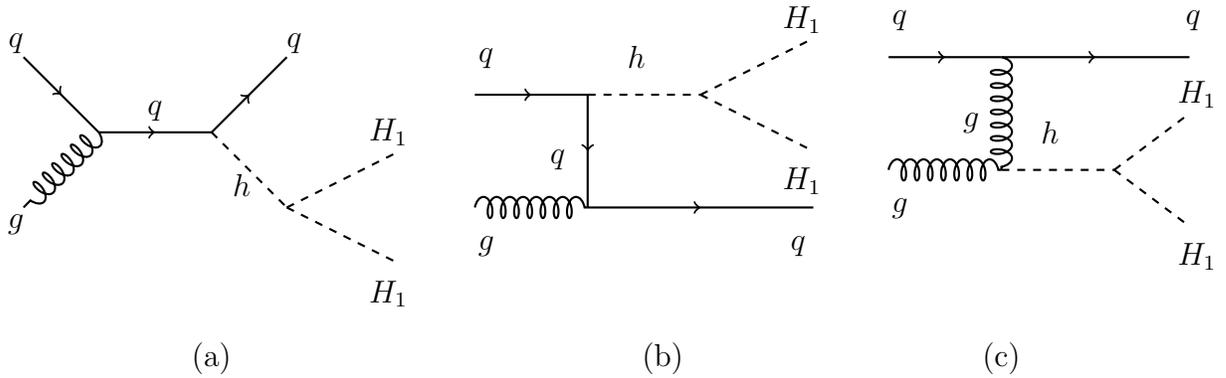
      \end{minipage}
      \end{minipage}

\subsection{VBF diagrams with dijet final states}

\begin{minipage}{0.9\linewidth}
\centering
\begin{minipage}{0.9\linewidth}
\centering
          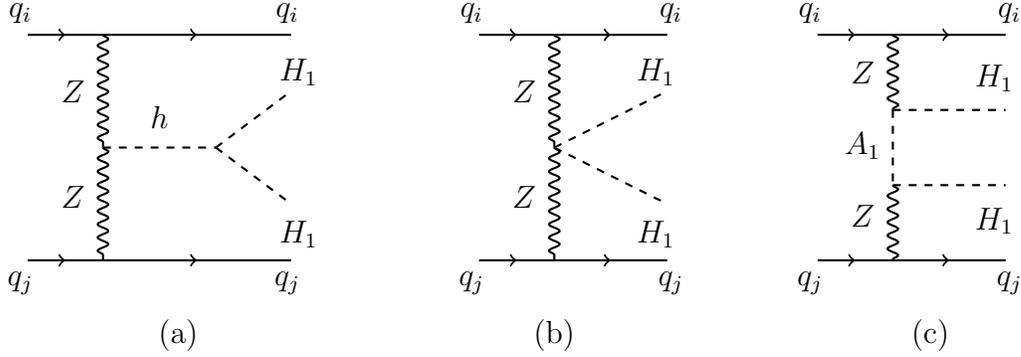
\begin{figure}[H]
%\centering          
             \begin{tikzpicture}[thick,scale=1]
\draw[particle] (0,0) -- node[black,above,xshift=-0.6cm,yshift=0.0cm] {$q_i$} (1,0);
\draw[photon] (1,0) -- node[black,above,xshift=-0.4cm,yshift=-0.3cm] {$Z$} (1,-1.5);
\draw[photon] (1,-1.5) -- node[black,above,xshift=-0.4cm,yshift=-0.2cm] {$Z$} (1,-3);
\draw[particle] (0,-3) -- node[black,above,xshift=-0.6cm,yshift=-0.6cm] {$q_j$} (1,-3);
\draw[dashed] (1,-1.5) -- node[black,above,yshift=0.1cm,xshift=-0.0cm] {$h$} (2.5,-1.5);
\draw[dashed] (2.5,-1.5) -- node[black,above,yshift=0.3cm,xshift=0.6cm] {$H_1$} (3.5,-0.75);
\draw[dashed] (2.5,-1.5) -- node[black,above,yshift=-1.1cm,xshift=0.6cm] {$H_1$} (3.5,-2.25);
\draw[particle] (1,0) -- node[black,above,xshift=1.2cm,yshift=0.0cm] {$q_i$} (3.5,0);
\draw[particle] (1,-3) -- node[black,above,xshift=1.2cm,yshift=-0.6cm] {$q_j$} (3.5,-3);

\node at (2,-4) {(a)};

%\quad
%\quad
\hspace{1cm}

\draw[particle] (5,0) -- node[black,above,xshift=-0.6cm,yshift=0.0cm] {$q_i$} (6,0);
\draw[photon] (6,0) -- node[black,above,xshift=-0.4cm,yshift=-0.3cm] {$Z$} (6,-1.5);
\draw[photon] (6,-1.5) -- node[black,above,xshift=-0.4cm,yshift=-0.2cm] {$Z$} (6,-3);
\draw[particle] (5,-3) -- node[black,above,xshift=-0.6cm,yshift=-0.6cm] {$q_j$} (6,-3);
\draw[particle] (6,0) -- node[black,above,xshift=0.8cm,yshift=0.0cm] {$q_i$} (7.5,0);
\draw[particle] (6,-3) -- node[black,above,xshift=0.8cm,yshift=-0.6cm] {$q_j$} (7.5,-3);
\draw[dashed] (6,-1.5) -- node[black,above,yshift=0.3cm,xshift=0.6cm] {$H_1$} (7.5,-0.75);
\draw[dashed] (6,-1.5) -- node[black,above,yshift=-1.1cm,xshift=0.6cm] {$H_1$} (7.5,-2.25);

\node at (6,-4) {(b)};

\hspace{1cm}

\draw[particle] (8.5,0) -- node[black,above,xshift=-0.6cm,yshift=0.0cm] {$q_i$} (9.5,0);
\draw[photon] (9.5,0) -- node[black,above,xshift=-0.4cm,yshift=-0.3cm] {$Z$} (9.5,-1);
\draw[dashed] (9.5,-1) -- node[black,above,xshift=-0.4cm,yshift=-0.3cm] {$A_1$} (9.5,-2);
\draw[photon] (9.5,-2) -- node[black,above,xshift=-0.4cm,yshift=-0.2cm] {$Z$} (9.5,-3);
\draw[particle] (8.5,-3) -- node[black,above,xshift=-0.6cm,yshift=-0.6cm] {$q_j$} (9.5,-3);
\draw[particle] (9.5,0) -- node[black,above,xshift=0.8cm,yshift=0.0cm] {$q_i$} (11,0);
\draw[particle] (9.5,-3) -- node[black,above,xshift=0.8cm,yshift=-0.6cm] {$q_j$} (11,-3);
\draw[dashed] (9.5,-1) -- node[black,above,yshift=0.1cm,xshift=0.6cm] {$H_1$} (11,-1);
\draw[dashed] (9.5,-2) -- node[black,above,yshift=-0.8cm,xshift=0.6cm] {$H_1$} (11,-2);

\node at (10,-4) {(c)};

%\draw\mypath;
\end{tikzpicture}
              \caption{Relevant VBF diagrams with dijet final states ($q_i q_j \to H_1 H_1 q_i q_j$) with only neutral intermediate gauge bosons, where $q=u,d$. Note that only one of the involved diagrams in this process depends on the value of the Higgs-DM coupling.}
\label{diag:nvbf}
          \end{figure}
      \end{minipage}
      \end{minipage}

\begin{minipage}{0.9\linewidth}
\centering
\begin{minipage}{0.9\linewidth}
\centering
          \begin{figure}[H]
%\centering          
             \begin{tikzpicture}[thick,scale=1.0]
\draw[particle] (0,0) -- node[black,above,xshift=-0.6cm,yshift=0.0cm] {$q_i$} (1,0);
\draw[photon] (1,0) -- node[black,above,xshift=-0.4cm,yshift=-0.3cm] {$W^+$} (1,-1.5);
\draw[photon] (1,-1.5) -- node[black,above,xshift=-0.4cm,yshift=-0.2cm] {$W^+$} (1,-3);
\draw[particle] (0,-3) -- node[black,above,xshift=-0.6cm,yshift=-0.6cm] {$q_j$} (1,-3);
\draw[dashed] (1,-1.5) -- node[black,above,yshift=0.1cm,xshift=-0.0cm] {$h$} (2.5,-1.5);
\draw[dashed] (2.5,-1.5) -- node[black,above,yshift=0.3cm,xshift=0.6cm] {$H_1$} (3.5,-0.75);
\draw[dashed] (2.5,-1.5) -- node[black,above,yshift=-1.1cm,xshift=0.6cm] {$H_1$} (3.5,-2.25);
\draw[particle] (1,0) -- node[black,above,xshift=1.2cm,yshift=0.0cm] {$q_k$} (3.5,0);
\draw[particle] (1,-3) -- node[black,above,xshift=1.2cm,yshift=-0.6cm] {$q_l$} (3.5,-3);

\node at (2,-4) {(a)};

%\quad
%\quad
\hspace{1cm}

\draw[particle] (5,0) -- node[black,above,xshift=-0.6cm,yshift=0.0cm] {$q_i$} (6,0);
\draw[photon] (6,0) -- node[black,above,xshift=-0.4cm,yshift=-0.3cm] {$W^+$} (6,-1.5);
\draw[photon] (6,-1.5) -- node[black,above,xshift=-0.4cm,yshift=-0.2cm] {$W^+$} (6,-3);
\draw[particle] (5,-3) -- node[black,above,xshift=-0.6cm,yshift=-0.6cm] {$q_j$} (6,-3);
\draw[particle] (6,0) -- node[black,above,xshift=0.8cm,yshift=0.0cm] {$q_k$} (7.5,0);
\draw[particle] (6,-3) -- node[black,above,xshift=0.8cm,yshift=-0.6cm] {$q_l$} (7.5,-3);
\draw[dashed] (6,-1.5) -- node[black,above,yshift=0.3cm,xshift=0.6cm] {$H_1$} (7.5,-0.75);
\draw[dashed] (6,-1.5) -- node[black,above,yshift=-1.1cm,xshift=0.6cm] {$H_1$} (7.5,-2.25);

\node at (6,-4) {(b)};

\hspace{1cm}

\draw[particle] (8.5,0) -- node[black,above,xshift=-0.6cm,yshift=0.0cm] {$q_i$} (9.5,0);
\draw[photon] (9.5,0) -- node[black,above,xshift=-0.4cm,yshift=-0.3cm] {$W^+$} (9.5,-1);
\draw[dashed] (9.5,-1) -- node[black,above,xshift=-0.4cm,yshift=-0.3cm] {$H^+_1$} (9.5,-2);
\draw[photon] (9.5,-2) -- node[black,above,xshift=-0.4cm,yshift=-0.2cm] {$W^+$} (9.5,-3);
\draw[particle] (8.5,-3) -- node[black,above,xshift=-0.6cm,yshift=-0.6cm] {$q_j$} (9.5,-3);
\draw[particle] (9.5,0) -- node[black,above,xshift=0.8cm,yshift=0.0cm] {$q_k$} (11,0);
\draw[particle] (9.5,-3) -- node[black,above,xshift=0.8cm,yshift=-0.6cm] {$q_l$} (11,-3);
\draw[dashed] (9.5,-1) -- node[black,above,yshift=0.1cm,xshift=0.6cm] {$H_1$} (11,-1);
\draw[dashed] (9.5,-2) -- node[black,above,yshift=-0.8cm,xshift=0.6cm] {$H_1$} (11,-2);

\node at (10,-4) {(c)};

%\draw\mypath;
\end{tikzpicture}
              \caption{Relevant VBF diagrams with dijet final states ($q_i q_j \to H_1 H_1 q_k q_l$) with only charged intermediate gauge bosons, where $q=u,d$. Note that only one of the involved diagrams in this process depends on the value of the Higgs-DM coupling.}
\label{diag:chvbf}
          \end{figure}
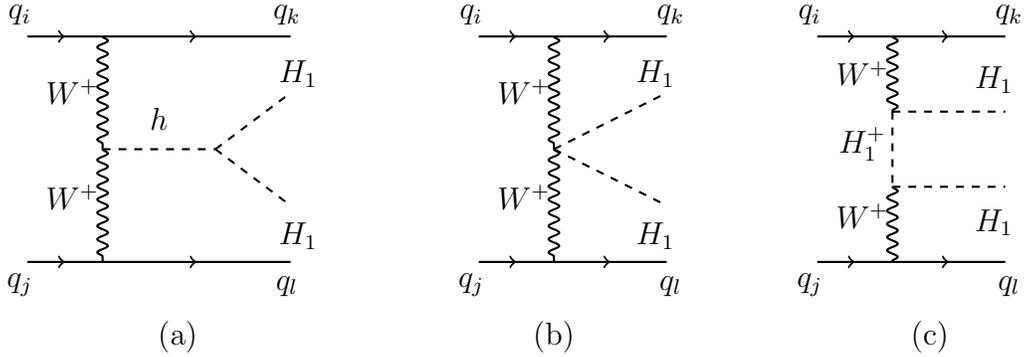
      \end{minipage}
      \end{minipage}

\subsection{HS diagrams with (on-shell) gauge boson final states}
\begin{minipage}{0.9\linewidth}
\centering
\begin{minipage}{0.9\linewidth}
\centering
\begin{figure}[H]
%\centering 
\begin{tikzpicture}[thick,scale=1.0]

\hspace{-1cm}

\draw[particle] (0,0) -- node[black,above,xshift=-0.6cm,yshift=0.4cm] {$q_i$} (1,-0.75);
\draw[antiparticle] (0,-1.5) -- node[black,above,yshift=-1.0cm,xshift=-0.6cm] {$\bar{q}_i$} (1,-0.75);
\draw[photon] (1,-0.75) -- node[black,above,xshift=0.0cm,yshift=0.0cm] {$Z$} (2,-0.75);
%\draw[dashed] (4,-1) -- node[black,above,yshift=-0.5cm,xshift=-0.2cm] {$\varphi$} (4,-2);
%\draw[particle] (4,-1) -- node[black,above,xshift=0.0cm,yshift=0.0cm] {$\chi_{b}$} (5,-1);
\draw[dashed] (2,-0.75) -- node[black,above,yshift=0.0cm,xshift=-0.0cm] {$h$} (3,-0.75);
\draw[dashed] (3,-0.75) -- node[black,above,yshift=0.1cm,xshift=0.8cm] {$H_1$} (4,-0);
\draw[dashed] (3,-0.75) -- node[black,above,yshift=-0.4cm,xshift=0.8cm] {$H_1$} (4,-1.5);
%\draw[photon] (2,-0.75) -- node[black,above,xshift=-0.45cm,yshift=-0.3cm] {$Z$} (3,-2.5);
%\draw[particle] (3,-2.5) -- node[black,above,xshift=0.9cm,yshift=-0.1cm] {$q_j$} (4.5,-1.75);
%\draw[antiparticle] (3,-2.5) -- node[black,above,yshift=-0.6cm,xshift=0.9cm] {$\bar{q}_j$} (4.5,-3.25);
\draw[photon] (2,-0.75) -- node[black,above,xshift=1.1cm,yshift=-1.5cm] {$Z^{*}$} (4,-2.5);

\node at (2,-4) {(a)};

\hspace{1cm}

\draw[particle] (5,0) -- node[black,above,xshift=-0.6cm,yshift=0.4cm] {$q$} (6,-0.75);
\draw[antiparticle] (5,-1.5) -- node[black,above,yshift=-1.0cm,xshift=-0.6cm] {$\bar{q}$} (6,-0.75);
\draw[photon] (6,-0.75) -- node[black,above,xshift=0.0cm,yshift=0.0cm] {$Z$} (7,-0.75);
%\draw[dashed] (4,-1) -- node[black,above,yshift=-0.5cm,xshift=-0.2cm] {$\varphi$} (4,-2);
%\draw[particle] (4,-1) -- node[black,above,xshift=0.0cm,yshift=0.0cm] {$\chi_{b}$} (5,-1);
%\draw[dashed] (7,-0.75) -- node[black,above,yshift=0.0cm,xshift=-0.0cm] {$h$} (8,-0.75);
\draw[dashed] (7,-0.75) -- node[black,above,yshift=0.1cm,xshift=1.2cm] {$H_1$} (8.5,-0);
\draw[dashed] (7,-0.75) -- node[black,above,yshift=-0.4cm,xshift=1.1cm] {$H_1$} (8.5,-1.5);
%\draw[photon] (7,-0.75) -- node[black,above,xshift=-0.45cm,yshift=-0.3cm] {$Z$} (8,-2.5);
%\draw[particle] (8,-2.5) -- node[black,above,xshift=0.7cm,yshift=-0.1cm] {$q$} (9,-1.75);
%\draw[antiparticle] (8,-2.5) -- node[black,above,xshift=0.6cm,yshift=-0.6cm] {$\bar{q}$} (9,-3.25);
\draw[photon] (7,-0.75) -- node[black,above,xshift=1.15cm,yshift=-1.5cm] {$Z^{*}$} (8.5,-2.5);

\node at (6,-4) {(b)};

\hspace{1cm}

\draw[particle] (10,0) -- node[black,above,xshift=-0.6cm,yshift=0.4cm] {$q$} (11,-0.75);
\draw[antiparticle] (10,-1.5) -- node[black,above,yshift=-1.0cm,xshift=-0.6cm] {$\bar{q}$} (11,-0.75);
\draw[photon] (11,-0.75) -- node[black,above,xshift=0.0cm,yshift=0.0cm] {$Z$} (12,-0.75);
\draw[dashed] (12,-0.75) -- node[black,above,yshift=0.1cm,xshift=0.8cm] {$H_1$} (13,-0);
\draw[dashed] (12,-0.75) -- node[black,above,yshift=-0.3cm,xshift=-0.3cm] {$A_1$} (12,-1.75);
\draw[dashed] (12,-1.75) -- node[black,above,yshift=0.1cm,xshift=0.8cm] {$H_1$} (13,-1);
%\draw[photon] (12,-1.75) -- node[black,above,xshift=-0.3cm,yshift=-0.3cm] {$Z$} (12,-2.75);
%\draw[particle] (12,-2.75) -- node[black,above,xshift=0.7cm,yshift=-0.1cm] {$q$} (13,-2);
%\draw[antiparticle] (12,-2.75) -- node[black,above,xshift=0.6cm,yshift=-0.6cm] {$\bar{q}$} (13,-3.5);
\draw[photon] (12,-1.75) -- node[black,above,xshift=0.6cm,yshift=-0.9cm] {$Z^{*}$} (13,-2.5);

\node at (11,-4) {(c)};

%\draw\mypath;
\end{tikzpicture}
              \caption{Relevant HS diagrams with (on-shell) neutral gauge boson final states ($q_i \bar{q_i} \to H_1 H_1 Z^{*}$) and only neutral intermediate gauge bosons, where $q=u,d$. Note that only one of the involved diagrams in this process depends on the value of the Higgs-DM coupling.}
\label{diag:nstr}
          \end{figure}
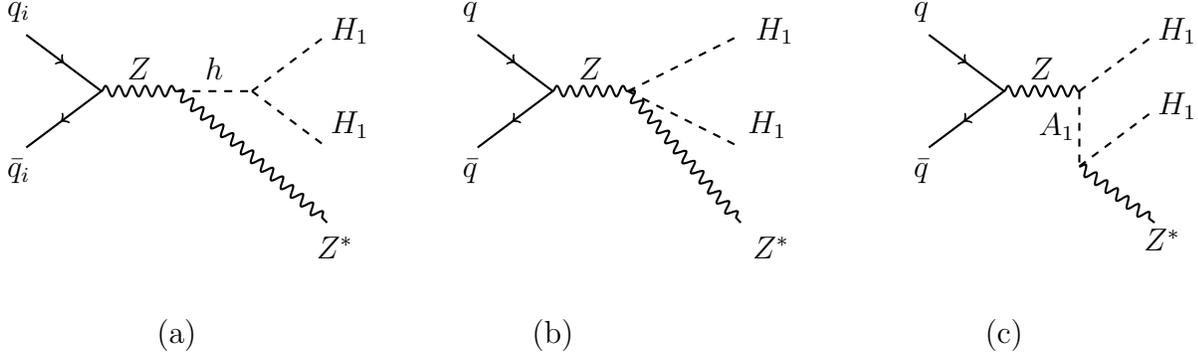
      \end{minipage}
      \end{minipage}

\begin{minipage}{0.9\linewidth}
\centering
\begin{minipage}{0.9\linewidth}
\centering
\begin{figure}[H]
%\centering 
 \begin{tikzpicture}[thick,scale=1.0]

\hspace{-1cm}

\draw[particle] (0,0) -- node[black,above,xshift=-0.6cm,yshift=0.4cm] {$q_i$} (1,-0.75);
\draw[antiparticle] (0,-1.5) -- node[black,above,yshift=-1.0cm,xshift=-0.6cm] {$\bar{q}_j$} (1,-0.75);
\draw[photon] (1,-0.75) -- node[black,above,xshift=0.0cm,yshift=0.0cm] {$W^+$} (2,-0.75);
%\draw[dashed] (4,-1) -- node[black,above,yshift=-0.5cm,xshift=-0.2cm] {$\varphi$} (4,-2);
%\draw[particle] (4,-1) -- node[black,above,xshift=0.0cm,yshift=0.0cm] {$\chi_{b}$} (5,-1);
\draw[dashed] (2,-0.75) -- node[black,above,yshift=0.0cm,xshift=-0.0cm] {$h$} (3,-0.75);
\draw[dashed] (3,-0.75) -- node[black,above,yshift=0.1cm,xshift=0.8cm] {$H_1$} (4,-0);
\draw[dashed] (3,-0.75) -- node[black,above,yshift=-0.4cm,xshift=0.8cm] {$H_1$} (4,-1.5);
\draw[photon] (2,-0.75) -- node[black,above,xshift=1.25cm,yshift=-1.5cm] {$W^{+*}$} (4,-2.5);
%\draw[particle] (3,-2.5) -- node[black,above,xshift=0.9cm,yshift=-0.1cm] {$q_k$} (4.5,-1.75);
%\draw[antiparticle] (3,-2.5) -- node[black,above,yshift=-0.6cm,xshift=0.9cm] {$\bar{q}_l$} (4.5,-3.25);

\node at (2,-4) {(a)};

\hspace{1cm}

\draw[particle] (5,0) -- node[black,above,xshift=-0.6cm,yshift=0.4cm] {$q_i$} (6,-0.75);
\draw[antiparticle] (5,-1.5) -- node[black,above,yshift=-1.0cm,xshift=-0.6cm] {$\bar{q}_j$} (6,-0.75);
\draw[photon] (6,-0.75) -- node[black,above,xshift=0.0cm,yshift=0.0cm] {$W^+$} (7,-0.75);
%\draw[dashed] (4,-1) -- node[black,above,yshift=-0.5cm,xshift=-0.2cm] {$\varphi$} (4,-2);
%\draw[particle] (4,-1) -- node[black,above,xshift=0.0cm,yshift=0.0cm] {$\chi_{b}$} (5,-1);
%\draw[dashed] (7,-0.75) -- node[black,above,yshift=0.0cm,xshift=-0.0cm] {$h$} (8,-0.75);
\draw[dashed] (7,-0.75) -- node[black,above,yshift=0.1cm,xshift=1.2cm] {$H_1$} (8.5,-0);
\draw[dashed] (7,-0.75) -- node[black,above,yshift=-0.4cm,xshift=1.1cm] {$H_1$} (8.5,-1.5);
\draw[photon] (7,-0.75) -- node[black,above,xshift=1.25cm,yshift=-1.5cm] {$W^{+*}$} (8.5,-2.5);
%\draw[particle] (8,-2.5) -- node[black,above,xshift=0.7cm,yshift=-0.1cm] {$q_k$} (9,-1.75);
%\draw[antiparticle] (8,-2.5) -- node[black,above,xshift=0.6cm,yshift=-0.6cm] {$\bar{q}_l$} (9,-3.25);

\node at (7,-4) {(b)};

\hspace{1cm}

\draw[particle] (10,0) -- node[black,above,xshift=-0.6cm,yshift=0.4cm] {$q_i$} (11,-0.75);
\draw[antiparticle] (10,-1.5) -- node[black,above,yshift=-1.0cm,xshift=-0.6cm] {$\bar{q}_j$} (11,-0.75);
\draw[photon] (11,-0.75) -- node[black,above,xshift=0.0cm,yshift=0.0cm] {$W^+$} (12,-0.75);
\draw[dashed] (12,-0.75) -- node[black,above,yshift=0.1cm,xshift=0.8cm] {$H_1$} (13,-0);
\draw[dashed] (12,-0.75) -- node[black,above,yshift=-0.3cm,xshift=-0.3cm] {$A_1$} (12,-1.75);
\draw[dashed] (12,-1.75) -- node[black,above,yshift=0.1cm,xshift=0.8cm] {$H_1$} (13,-1);
\draw[photon] (12,-1.75) -- node[black,above,xshift=0.6cm,yshift=-0.9cm] {$W^{+*}$} (13,-2.5);
%\draw[particle] (12,-2.75) -- node[black,above,xshift=0.7cm,yshift=-0.1cm] {$q_k$} (13,-2);
%\draw[antiparticle] (12,-2.75) -- node[black,above,xshift=0.6cm,yshift=-0.6cm] {$\bar{q}_l$} (13,-3.5);

\node at (11,-4) {(c)};

%\draw\mypath;
\end{tikzpicture}
\caption{Relevant HS diagrams with (on-shell) charged gauge boson final states ($q_i \bar{q_j} \to H_1 H_1 W^{*+}$) and only charged intermediate gauge bosons, where $q=u,d$. Note that only one of the involved diagrams in this process depends on the value of the Higgs-DM coupling.}
\label{diag:chstr}
          \end{figure}
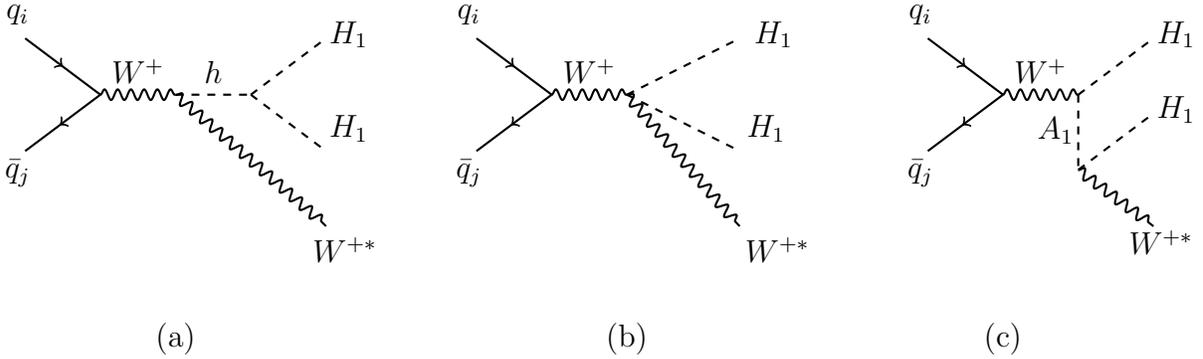
      \end{minipage}
      \end{minipage}


\begin{thebibliography}{99}
\bibitem{Aad:2012tfa}
{ATLAS} Collaboration, %G.~Aad {\em et.~al.}, 
%{\it {Observation of a new
  %particle in the search for the Standard Model Higgs boson with the ATLAS
  %detector at the LHC}},  
{\em Phys.Lett.} {\bf B716} (2012) 1.%--29,
%  [\href{http://xxx.lanl.gov/abs/1207.7214}{{\tt arXiv:1207.7214}}].

\bibitem{Chatrchyan:2012ufa}
{CMS} Collaboration, %S.~Chatrchyan {\em et.~al.}, %{\it {Observation of a
  %new boson at a mass of 125 GeV with the CMS experiment at the LHC}},  
{\em
  Phys.Lett.} {\bf B716} (2012) 30.%--61,
%  [\href{http://xxx.lanl.gov/abs/1207.7235}{{\tt arXiv:1207.7235}}].

%\cite{Ade:2015xua}
\bibitem{Ade:2015xua}
  P.~A.~R.~Ade {\it et al.} [Planck Collaboration],
  %``Planck 2015 results. XIII. Cosmological parameters,''
  arXiv:1502.01589 [astro-ph.CO].
  %%CITATION = ARXIV:1502.01589;%%
  %464 citations counted in INSPIRE as of 25 Jul 2015
  
\bibitem{Jungman:1995df}
G.~Jungman, M.~Kamionkowski and K.~Griest, %{\it {Supersymmetric dark matter}},
   {\em Phys.Rept.} {\bf 267} (1996) 195.%--373,
%  [\href{http://xxx.lanl.gov/abs/hep-ph/9506380}{{\tt hep-ph/9506380}}].

\bibitem{Bertone:2004pz}
G.~Bertone, D.~Hooper and J.~Silk, %{\it {Particle dark matter: Evidence,
%  candidates and constraints}},  
{\em Phys.Rept.} {\bf 405} (2005) 279.%--390,
%  [\href{http://xxx.lanl.gov/abs/hep-ph/0404175}{{\tt hep-ph/0404175}}].

\bibitem{Bergstrom:2000pn}
L.~Bergstrom, %{\it {Nonbaryonic dark matter: Observational evidence and
%  detection methods}}, 
 {\em Rept.Prog.Phys.} {\bf 63} (2000) 793.
%  [\href{http://xxx.lanl.gov/abs/hep-ph/0002126}{{\tt hep-ph/0002126}}].



\bibitem{Deshpande:1977rw}
N.~G. Deshpande and E.~Ma, 
%``{Pattern of Symmetry Breaking with Two Higgs  Doublets},'' 
{\em Phys.Rev.} {\bf D18} (1978) 2574.

%\cite{King:1984vr}
\bibitem{King:1984vr}
  S.~F.~King,
  %``Monojets From $Z$ Decay Without Extra Neutrinos or Higgs,''
  Phys.\ Rev.\ Lett.\  {\bf 54} (1985) 528.
  %%CITATION = PRLTA,54,528;%%
  %12 citations counted in INSPIRE as of 25 Jul 2015


%\bibitem{Gustafsson:2007pc}
%M.~Gustafsson, E.~Lundstrom, L.~Bergstrom, and J.~Edsjo, ``Significant gamma
%  lines from inert higgs dark matter,'' 2007, astro-ph/0703512.


\bibitem{Ma:2006km}
E.~Ma,
% {\it {Verifiable radiative seesaw mechanism of neutrino mass and dark  matter}},  
{\em Phys.Rev.} {\bf D73} (2006) 077301,
  %[\href{http://xxx.lanl.gov/abs/hep-ph/0601225}{{\tt hep-ph/0601225}}].


\bibitem{Barbieri:2006dq}
R.~Barbieri, L.~J. Hall and V.~S. Rychkov, 
%``Improved naturalness with a heavy  higgs: An alternative road to lhc physics,'' 
{\em Phys. Rev.} {\bf D74} (2006) 015007. %hep-ph/0603188.
  
\bibitem{LopezHonorez:2006gr}
L.~Lopez~Honorez, E.~Nezri, J.~F. Oliver and M.~H.~G. Tytgat, 
%``The inert  doublet model: An archetype for dark matter,'' 
{\em JCAP} {\bf 0702} (2007) 028. % hep-ph/0612275.



\bibitem{Krawczyk:2013jta}
  M.~Krawczyk, D.~Sokolowska, P.~Swaczyna and B.~Swiezewska,
  %``Constraining Inert Dark Matter by $R_{\gamma\gamma}$ and WMAP data,''
  JHEP {\bf 1309} (2013) 055.
%  [arXiv:1305.6266 [hep-ph]].
  %%CITATION = ARXIV:1305.6266;%%
  %33 citations counted in INSPIRE as of 10 juil. 2015


%\cite{Arhrib:2013ela}
\bibitem{Arhrib:2013ela}
  A.~Arhrib, Y.~L.~S.~Tsai, Q.~Yuan and T.~C.~Yuan,
  %``An Updated Analysis of Inert Higgs Doublet Model in light of the Recent Results from LUX, PLANCK, AMS-02 and LHC,''
  JCAP {\bf 1406} (2014) 030
  [arXiv:1310.0358 [hep-ph]].
  %%CITATION = ARXIV:1310.0358;%%
  %35 citations counted in INSPIRE as of 29 juil. 2015


%\cite{Keus:2014jha}
\bibitem{Keus:2014jha}
  V.~Keus, S.~F.~King, S.~Moretti and D.~Sokolowska,
  %``Dark Matter with Two Inert Doublets plus One Higgs Doublet,''
  JHEP {\bf 1411} (2014) 016.
%  [arXiv:1407.7859 [hep-ph]].
  %%CITATION = ARXIV:1407.7859;%%
  %9 citations counted in INSPIRE as of 21 Jun 2015

%\cite{Ivanov:2011ae}
\bibitem{Ivanov:2011ae} 
  I.~P.~Ivanov, V.~Keus and E.~Vdovin,
  %``Abelian symmetries in multi-Higgs-doublet models,''
  J.\ Phys.\ A {\bf 45} (2012) 215201.
  %[arXiv:1112.1660 [math-ph]].
  %%CITATION = ARXIV:1112.1660;%%
  %11 citations counted in INSPIRE as of 28 Apr 2014
  
%\cite{Moretti:2015cwa}
\bibitem{Moretti:2015cwa}
  S.~Moretti and K.~Yagyu,
  %``Constraints on Parameter Space from Perturbative Unitarity in Models with Three Scalar Doublets,''
  Phys.\ Rev.\ D {\bf 91} (2015) 055022.
%  [arXiv:1501.06544 [hep-ph]].
  %%CITATION = ARXIV:1501.06544;%%
  %1 citations counted in INSPIRE as of 27 juil. 2015
  

%\cite{Akerib:2013tjd}
\bibitem{Akerib:2013tjd}
  D.~S.~Akerib {\it et al.} [LUX Collaboration],
  %``First results from the LUX dark matter experiment at the Sanford Underground Research Facility,''
  Phys.\ Rev.\ Lett.\  {\bf 112} (2014) 091303
  [arXiv:1310.8214 [astro-ph.CO]].
  %%CITATION = ARXIV:1310.8214;%%
  %832 citations counted in INSPIRE as of 11 juil. 2015
  
\bibitem{Aprile:2012zx}
  E.~Aprile [XENON1T Collaboration],
  %``The XENON1T Dark Matter Search Experiment,''
  Springer Proc.\ Phys.\  {\bf 148} (2013) 93
  [arXiv:1206.6288 [astro-ph.IM]].
  %%CITATION = ARXIV:1206.6288;%%
  %257 citations counted in INSPIRE as of 11 Jul 2015  

  \bibitem{Anderson:2011bi}
  A.~J.~Anderson, J.~M.~Conrad, E.~Figueroa-Feliciano, K.~Scholberg and J.~Spitz,
  %``Coherent Neutrino Scattering in Dark Matter Detectors,''
  Phys.\ Rev.\ D {\bf 84} (2011) 013008
  [arXiv:1103.4894 [hep-ph]].
  %%CITATION = ARXIV:1103.4894;%%
  %29 citations counted in INSPIRE as of 11 juil. 2015



%\cite{LopezHonorez:2010tb}
\bibitem{LopezHonorez:2010tb}
  L.~Lopez Honorez and C.~E.~Yaguna,
  %``A new viable region of the inert doublet model,''
  JCAP {\bf 1101} (2011) 002.
%  [arXiv:1011.1411 [hep-ph]].
  %%CITATION = ARXIV:1011.1411;%%
  %69 citations counted in INSPIRE as of 24 Apr 2015

%\cite{Keus:2014isa}
\bibitem{Keus:2014isa}
  V.~Keus, S.~F.~King and S.~Moretti,
  %``Phenomenology of the inert ( 2+1 ) and ( 4+2 ) Higgs doublet models,''
  Phys.\ Rev.\ D {\bf 90} (2014)  075015.
%  [arXiv:1408.0796 [hep-ph]].
  %%CITATION = ARXIV:1408.0796;%%
  %5 citations counted in INSPIRE as of 07 Jul 2015
 

%\cite{Semenov:2014rea}
\bibitem{Semenov:2014rea}
  A.~Semenov,
  %``LanHEP - a package for automatic generation of Feynman rules from the Lagrangian. Updated version 3.2,''
  arXiv:1412.5016 [physics.comp-ph].
  %%CITATION = ARXIV:1412.5016;%%
  
%\cite{Belyaev:2012qa}
\bibitem{Belyaev:2012qa}
  A.~Belyaev, N.~D.~Christensen and A.~Pukhov,
  %``CalcHEP 3.4 for collider physics within and beyond the Standard Model,''
  Comput.\ Phys.\ Commun.\  {\bf 184} (2013) 1729.
%  [arXiv:1207.6082 [hep-ph]].
  %%CITATION = ARXIV:1207.6082;%%
  %238 citations counted in INSPIRE as of 07 juil. 2015

  
%\cite{Garcia-Cely:2013zga}
\bibitem{Garcia-Cely:2013zga}
  C.~Garcia-Cely and A.~Ibarra,
  %``Novel Gamma-ray Spectral Features in the Inert Doublet Model,''
  JCAP {\bf 1309} (2013) 025.
 % [arXiv:1306.4681 [hep-ph]].
  %%CITATION = ARXIV:1306.4681;%%
  %12 citations counted in INSPIRE as of 24 Apr 2015
  
\end{thebibliography}
\end{document}